\documentclass[prl,aps,epsfig,superscriptaddress,twocolumn]{revtex4-1}
\setcitestyle{super}
\usepackage{amsfonts,amssymb,amscd,amsthm}
\usepackage{graphicx}
\usepackage{mathrsfs}
\usepackage[intlimits]{amsmath}
\usepackage[colorlinks, citecolor=red]{hyperref}
\usepackage{etoolbox}
\usepackage{upgreek}
\begin{document}

\title{A Site-Resolved 2D Quantum Simulator with Hundreds of Trapped Ions}

\author{S.-A. Guo}
\affiliation{Center for Quantum Information, Institute for Interdisciplinary Information Sciences, Tsinghua University, Beijing 100084, PR China}

\author{Y.-K. Wu}
\affiliation{Center for Quantum Information, Institute for Interdisciplinary Information Sciences, Tsinghua University, Beijing 100084, PR China}
\affiliation{Hefei National Laboratory, Hefei 230088, PR China}

\author{J. Ye}
\affiliation{Center for Quantum Information, Institute for Interdisciplinary Information Sciences, Tsinghua University, Beijing 100084, PR China}

\author{L. Zhang}
\affiliation{Center for Quantum Information, Institute for Interdisciplinary Information Sciences, Tsinghua University, Beijing 100084, PR China}

\author{W.-Q. Lian}
\affiliation{HYQ Co., Ltd., Beijing 100176, PR China}

\author{R. Yao}
\affiliation{HYQ Co., Ltd., Beijing 100176, PR China}

\author{Y. Wang}
\affiliation{Center for Quantum Information, Institute for Interdisciplinary Information Sciences, Tsinghua University, Beijing 100084, PR China}
\affiliation{HYQ Co., Ltd., Beijing 100176, PR China}

\author{R.-Y. Yan}
\affiliation{Center for Quantum Information, Institute for Interdisciplinary Information Sciences, Tsinghua University, Beijing 100084, PR China}

\author{Y.-J. Yi}
\affiliation{Center for Quantum Information, Institute for Interdisciplinary Information Sciences, Tsinghua University, Beijing 100084, PR China}

\author{Y.-L. Xu}
\affiliation{Center for Quantum Information, Institute for Interdisciplinary Information Sciences, Tsinghua University, Beijing 100084, PR China}

\author{B.-W. Li}
\affiliation{HYQ Co., Ltd., Beijing 100176, PR China}

\author{Y.-H. Hou}
\affiliation{Center for Quantum Information, Institute for Interdisciplinary Information Sciences, Tsinghua University, Beijing 100084, PR China}

\author{Y.-Z. Xu}
\affiliation{Center for Quantum Information, Institute for Interdisciplinary Information Sciences, Tsinghua University, Beijing 100084, PR China}

\author{W.-X. Guo}
\affiliation{HYQ Co., Ltd., Beijing 100176, PR China}

\author{C. Zhang}
\affiliation{Center for Quantum Information, Institute for Interdisciplinary Information Sciences, Tsinghua University, Beijing 100084, PR China}

\author{B.-X. Qi}
\affiliation{Center for Quantum Information, Institute for Interdisciplinary Information Sciences, Tsinghua University, Beijing 100084, PR China}

\author{Z.-C. Zhou}
\affiliation{Center for Quantum Information, Institute for Interdisciplinary Information Sciences, Tsinghua University, Beijing 100084, PR China}
\affiliation{Hefei National Laboratory, Hefei 230088, PR China}

\author{L. He}
\affiliation{Center for Quantum Information, Institute for Interdisciplinary Information Sciences, Tsinghua University, Beijing 100084, PR China}
\affiliation{Hefei National Laboratory, Hefei 230088, PR China}

\author{L.-M. Duan}
\email{lmduan@tsinghua.edu.cn}
\affiliation{Center for Quantum Information, Institute for Interdisciplinary Information Sciences, Tsinghua University, Beijing 100084, PR China}
\affiliation{Hefei National Laboratory, Hefei 230088, PR China}
\affiliation{New Cornerstone Science Laboratory, Beijing 100084, PR China}
\maketitle

\textbf{A large qubit capacity and an individual readout capability are two crucial requirements for large-scale quantum computing and simulation \cite{DiVincenzo2000}. As one of the leading physical platforms for quantum information processing, the ion trap has achieved quantum simulation of tens of ions with site-resolved readout in 1D Paul trap \cite{zhang2017Observation,joshi2023simulation,PRXQuantum.4.010302}, and that of hundreds of ions with global observables in 2D Penning trap \cite{britton2012engineered,bohnet2016quantum}. However, integrating these two features into a single system is still very challenging. Here we report the stable trapping of $512$ ions in a 2D Wigner crystal and the sideband cooling of their transverse motion. We demonstrate the quantum simulation of long-range quantum Ising models with tunable coupling strengths and patterns, with or without frustration, using $300$ ions. Enabled by the site resolution in the single-shot measurement, we observe rich spatial correlation patterns in the quasi-adiabatically prepared ground states, which allows us to verify quantum simulation results by comparing with the calculated collective phonon modes and with classical simulated annealing. We further probe the quench dynamics of the Ising model in a transverse field to demonstrate quantum sampling tasks. Our work paves the way for simulating classically intractable quantum dynamics and for running NISQ algorithms \cite{RevModPhys.86.153,RevModPhys.94.015004} using 2D ion trap quantum simulators.}

Quantum computation and quantum simulation have entered an era of hundreds of qubits \cite{ebadi2021quantum,kim2023evidence,bohnet2016quantum}, with complicated computational tasks beyond the reach of current classical computers being demonstrated \cite{arute2019quantum,jiuzhang2020,PhysRevLett.127.180501,madsen2022quantum}. To further extend the application of the noisy intermediate-scale quantum (NISQ) devices on practical and classically intractable problems, the quantum simulation of many-body dynamics \cite{RevModPhys.86.153} and the NISQ algorithms \cite{RevModPhys.94.015004} like quantum annealing \cite{Hauke_2020} and variational quantum algorithms \cite{cerezo2021variational} have attracted wide research interest. Apart from a large qubit number, a critical requirement for these applications is the ability to read out individual qubit states in a single shot \cite{DiVincenzo2000}, thus allowing, e.g., the measurement of qubits' spatial correlation under quantum dynamics, or the evaluation of many-body objective functions in an optimization problem.

As one of the leading quantum computing platforms, the ion trap has demonstrated quantum simulation of up to $61$ qubits in a 1D Paul trap with individual detection \cite{zhang2017Observation,joshi2023simulation,PRXQuantum.4.010302}. To further scale up the qubit number, one plausible way is to trap the ions in a 2D crystal. To date, 2D crystals of up to $150$ ions under Doppler cooling \cite{Szymanski2012crystal,Xie_2021}, of about $50$ ions using two-tone laser cooling \cite{PhysRevA.105.023101}, and of about $100$ ions under EIT cooling \cite{PRXQuantum.4.020317} have been reported, and the global quantum manipulation and individual detection have also been demonstrated with up to $10$ ions in 2D \cite{qiao2022observing}. Incidentally, 2D micro-trap arrays have also been achieved in the small scale with large inter-site distances and weak coupling strength \cite{Hensinger2014array,PhysRevLett.123.213605,blatt2020_2d}, and 2D junctions for the quantum charge-coupled device (QCCD) architecture \cite{wineland1998experimental,kielpinski2002architecture} have also been realized as a plausible way to scale up the system
\cite{Amini_2010,PhysRevA.89.062308,PhysRevLett.130.173202}.
On the other hand, 2D ion crystals are native in Penning trap with quantum simulation of about $200$ ions already being achieved \cite{britton2012engineered,bohnet2016quantum}. However, due to the fast rotation of the ion crystal in a Penning trap, the individual detection of qubit states remains an experimental challenge. Although spatial and temporal resolved imaging techniques have been used to count the ion number \cite{britton2012engineered,bohnet2016quantum}, observables for quantum simulation in Penning trap are still limited to be global.

\begin{figure*}[!tbp]
	\centering
	\includegraphics[width=0.9\linewidth]{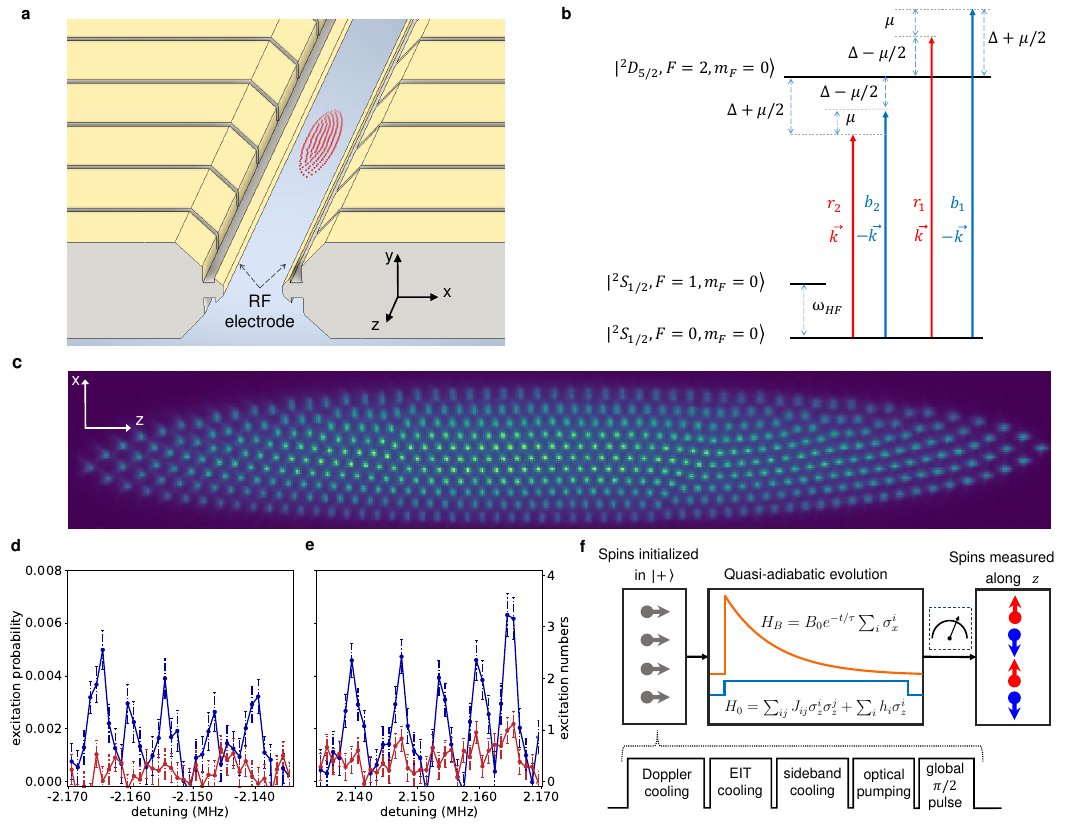}
	\caption { \textbf{Experimental setup and 2D ion crystals.} \textbf{a}, An illustration of our monolithic 3D ion trap at cryogenic temperature. The 2D ion crystal locates on the $xz$ plane and its size is not to scale. \textbf{b}, Relevant energy levels of the ${}^{171}\mathrm{Yb}^+$ ion. The qubit is encoded in the $S_{1/2}$ hyperfine levels $|0\rangle\equiv |F=0,m_F=0\rangle$ and $|1\rangle\equiv |F=1,m_F=0\rangle$, and can be rotated by a resonant global microwave. Counter-propagating off-resonant $411\,$nm laser beams are used to generate spin-dependent forces on the ions, which further lead to the effective Ising coupling when the phonon states are adiabatically eliminated. Two pairs of frequency components are placed on the two sides of the $S$-$D$ transition with detuning $\pm(\Delta\pm \mu/2)$ to create a beat note of $\mu$ while compensating their time-independent AC Stark shift. \textbf{c}, The image of a 2D ion crystal with $N=512$ ions. \textbf{d}, The spectrum of the red motional sideband of the transverse phonon modes and \textbf{e}, that of the blue motional sideband under Doppler cooling (blue) and sideband cooling (red). Here we only show the highest five modes including the center-of-mass (COM) mode, while the complete spectra can be found in Supplementary Information. \textbf{f}, The experimental sequence to quasi-adiabatically prepare the ground state of an Ising model Hamiltonian. We initialize all spins in $|0\rangle$ after laser cooling and optical pumping, and then rotate them to $|+\rangle$ by a global microwave SK1 composite $\pi/2$ pulse. Then we turn on the Ising model Hamiltonian $H_0$ via the global $411\,$nm laser and the transverse field $H_B$ via the global microwave simultaneously, and we quench the strength of the transverse field from $B_0\gg J_0$ following an exponential path, where $J_0=\frac{1}{N}\sum_{i\ne j} J_{ij}$ is the Kac normalized coupling strength.}\label{fig1}
\end{figure*}

Here, we report the stable trapping of a 2D ion crystal of $512$ ${}^{171}\mathrm{Yb}^+$ ions with its relevant transverse modes (perpendicular to the ion plane) EIT and sideband cooled to below one phonon per mode. We have exact control of the ion numbers in the crystal and can set it to any desired value. We further use the spin-dependent AC Stark shift of $411\,$nm laser to generate long-range Ising coupling, and demonstrate the quantum simulation of the long-range quantum Ising model using $300$ ions with tunable coupling patterns.
Note that comparable qubit numbers have also been achieved for neutral atoms \cite{ebadi2021quantum} and for superconducting circuits \cite{kim2023evidence} with individual-qubit resolution. In these systems, qubits mainly possess short-range nearest-neighbor interactions, while trapped ions naturally host long-range spin interactions with tunable coupling range and patterns. Due to this unprecedented control of coupling patterns, rich exotic spatial correlations can be expected. In particular, by tuning the laser close to individual phonon sidebands and quasi-adiabatically preparing the ground states via slow ramping of parameters, we observe various spin-spin spatial correlation patterns in which the collective oscillation modes of the ions are imprinted. We further demonstrate quantum simulations that are challenging for classical computers by simultaneously coupling to multiple phonon modes to obtain frustrated Ising coupling, or by probing the quench dynamics of a transverse-field Ising model and sampling from the final quantum states.

\section{Stable trapping of 2D ion crystals}
We use a monolithic 3D ion trap at the temperature of $T=6.1\,$K to hold the large 2D ion crystal. As discussed in detail in Sec. II of Supplementary Information, the cryogenic temperature is crucial for suppressing the collision influence with the background gas molecules and for the stability of the 2D crystal. By adjusting the voltages on the $4\times 7$ DC electrodes, we shape the crystal of $N=512$ ions into roughly an ellipse with 11 rows, while minimizing the micromotion perpendicular to the plane, as shown in Fig.~\ref{fig1}. The average distance between ions is about $a\approx 4\,\mu$m. Global $370\,$nm laser beams for Doppler cooling, EIT cooling, optical pumping and state detection are shined along the $yz$ plane so as to be immune to the micromotion of the ions in the $x$ direction. Owing to the coupling of the $x$ and $z$ directions in the in-plane motion, such cooling beams are sufficient to cool down the $x$ oscillation modes as well.

We further use a narrow-band global $411\,$nm laser in the $y$ direction to provide sideband cooling for the transverse modes. For $N=512$ ions under the transverse trap frequency $\omega_y=2\pi\times 2.164\,$MHz, the spectrum of the transverse modes spans about $2\pi\times 1.25\,$MHz below the center-of-mass (COM) mode. We perform sideband cooling to the highest $2\pi\times 150\,$kHz range by cooling $10$ frequencies, which covers all the phonon modes relevant to the following quantum simulation experiments. The red and blue motional sideband spectra for the highest $5$ modes are shown in Fig.~\ref{fig1}\textbf{d} and \textbf{e} under Doppler cooling (blue) and sideband cooling (red). To estimate the phonon number from these spectra, we maintain a low excitation rate for each phonon mode \cite{PhysRevLett.129.140501}, which results in the sensitivity to the state-preparation-and-measurement (SPAM) errors and the statistical fluctuation. Nevertheless, the current results allow us to bound the average phonon number to be below one phonon per mode, which is sufficient for our later quantum simulation of the spin model \cite{PhysRevLett.82.1971,PhysRevLett.103.120502}. More details about the setup, the daily operations and the estimation of the average phonon number can be found in Supplementary Information.

\begin{figure*}[!tbp]
	\centering
	\includegraphics[width=0.9\linewidth]{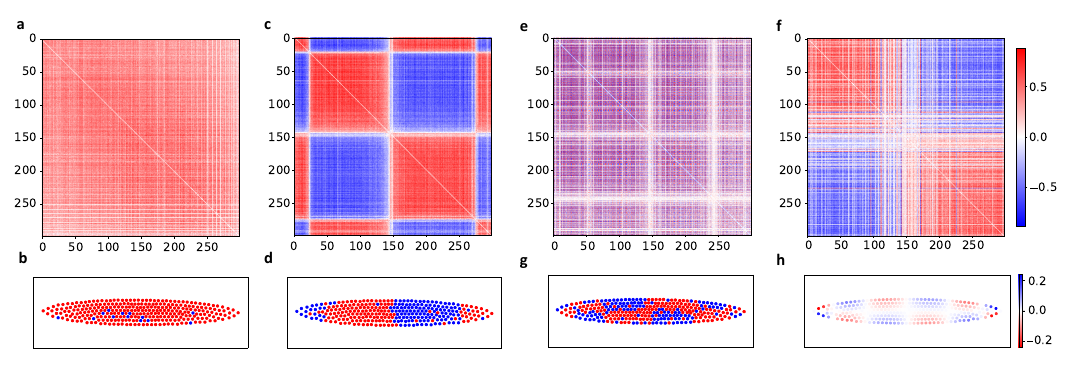}
	\caption {\textbf{Spatial correlation patterns in quasi-adiabatically prepared ground states for $N=300$ qubits.} \textbf{a}, The covariance matrix between all the ion pairs $C_{ij}\equiv \langle \sigma_z^i\sigma_z^j\rangle-\langle\sigma_z^i\rangle\langle\sigma_z^j\rangle$ (with the diagonal terms set to zero) when the beat note $\mu$ of the $411\,$nm laser is detuned by $\delta=\mu-\omega_1=2\pi\times 4\,$kHz above the COM mode $\omega_1=\omega_y=2\pi\times 2.140\,$MHz. The ions are labelled in ascending order of their $z$ coordinates. All the correlations are averaged over 100 samples. \textbf{b}, Typical single-shot measurement outcome. The blue and the red dots represent $|0\rangle$ and $|1\rangle$, respectively. Another typical pattern with most ions in $|0\rangle$ is not shown. Each ion can be resolved individually, and the overlap in the plot is due to the finite size of the markers. Small deviation from the ferromagnetic ground state comes from the nonadiabatic excitation and decoherence during the quench, as well as the detection error of the camera. \textbf{c}, \textbf{d}, Similar plots when the beat note $\mu$ of the $411\,$nm laser is detuned by $\delta=\mu-\omega_4=2\pi\times 1\,$kHz above the fourth highest phonon mode $\omega_4=\omega_y-2\pi\times 24.0\,$kHz.
\textbf{e}, Similar covariance matrix when the beat note is detuned by $\delta=\mu-\omega_{19}=2\pi\times 1\,$kHz above the 19th highest phonon mode $\omega_{19}=\omega_y-2\pi\times 139.2\,$kHz, with the ions labelled in ascending order of their $z$ coordinates. Since the correlations are oscillating both along the $x$ and the $z$ axes, when flattened into one axis, it is difficult to see the contiguous domains. \textbf{f}, The same covariance matrix as \textbf{e}, but with the ions rearranged in descending order of the mode coefficient $b_{i,19}$ as shown in \textbf{h}. \textbf{g}, Typical single-shot measurement result corresponding to \textbf{e}. \textbf{h}, Theoretically computed phonon mode structure $b_{i,19}$ for the 19th mode assuming harmonic trap potentials in three spatial directions.
}\label{fig2}
\end{figure*}

\section{Quantum simulation of long-range Ising model}
We further perform quantum simulation on a crystal of $N=300$ ions, which supports higher Ising coupling strength through narrower $411\,$nm global beams. As shown in Fig.~\ref{fig1}\textbf{b}, the Ising model Hamiltonian is achieved by the spin-dependent force from counter-propagating off-resonant $411\,$nm laser beams \cite{PhysRevA.103.012603} with the phonon states adiabatically eliminated. Here because the laser beams only couple $|0\rangle\equiv |S_{1/2},F=0,m_F=0\rangle$ to the $D_{5/2}$ levels but not $|1\rangle\equiv |S_{1/2},F=1,m_F=0\rangle$, the Ising Hamiltonian takes the form of

\begin{align}
H_0 =& \sum_{ij} J_{ij} (I+\sigma_z^i) (I+\sigma_z^j) \nonumber\\
\equiv & \sum_{i\ne j} J_{ij} \sigma_z^i \sigma_z^j + \sum_i h_i \sigma_z^i, \label{eq1}
\end{align}
where we have dropped an irrelevant constant. Apart from when coupling to the COM mode, the longitudinal field $h_i \equiv 2\sum_j J_{ij}$ is typically small (see Supplementary Information for details), and can further be compensated by a time-independent AC Stark shift of the $411\,$nm laser.

A transverse field $H_B=B\sum_i \sigma_x^i$ can be realized by a global microwave resonant to the qubit frequency. As shown in Fig.~\ref{fig1}\textbf{f}, we initialize all the qubits in $|+\rangle\equiv(|0\rangle+|1\rangle)/\sqrt{2}$ (prepared by the SK1 composite pulse to suppress the microwave inhomogeneity) and slowly ramp down the transverse field following an exponential path $B(t)=B_0e^{-t/\tau}$. By choosing $B_0> 50J_0$ where $J_0\equiv\frac{1}{N}\sum_{i\ne j} J_{ij}$ is the Kac normalized coupling strength, we start from the highest eigenstate of the Hamiltonian $H(t)=H_0+H_B(t)$. We further choose a total evolution time $T>5\tau$, and expect the final state to be close to the highest excited state of $H_0$, namely the ground state of $-H_0$.

Ideally when $h_i=0$, the system possesses $Z_2$ symmetry and we expect the final state to be invariant under the flip of all the spins. This shall give us large spin-spin correlation $C_{ij}\equiv \langle \sigma_z^i\sigma_z^j\rangle-\langle\sigma_z^i\rangle\langle\sigma_z^j\rangle=\langle \sigma_z^i\sigma_z^j\rangle$ in the prepared ground state. Even with the small uncompensated $h_i$'s, the pattern in the spatial correlation still survives, as can be seen in Fig.~\ref{fig2}. Specifically, if we couple dominantly to the COM mode with a positive detuning, it is well-known that $-H_0$ is an all-to-all coupled ferromagnetic Ising model whose ground state shows long-range correlation $C_{ij}=1$. This behavior can be observed in Fig.~\ref{fig2}\textbf{a} where we obtain positive correlations over almost all the ion pairs. The deviation from the ideal value of one can come from the nonadiabatic excitation or decoherence during the slow quench, and the state detection error for individual qubits (we use electron shelving to suppress the detection error to be below $1\%$ \cite{Roman2020,edmunds2020scalable,yang2022realizing}), as is evident from Fig.~\ref{fig2}\textbf{b} for the typical single-shot measurement outcomes. Besides, a reduced correlation can arise from a weak uncompensated longitudinal field, which will prefer one ferromagnetic ground state to the other, and thus increases the $\langle\sigma_z^i\rangle\langle\sigma_z^j\rangle$ term. Similarly, in Fig.~\ref{fig2}\textbf{c} and \textbf{d} we couple dominantly to the fourth highest phonon mode with a positive detuning, and observe a staggered spatial pattern.

Actually, when coupled to a single mode $k$, the ground state can be solved analytically and is governed by the structure of the phonon mode. This can be understood by factoring the Hamiltonian in Eq.~(\ref{eq1}) after the suppressing of the longitudinal field as
\begin{equation}
H_0 =\frac{1}{16\delta_k}\left(\sum_i\eta_k b_{ik}\Omega_i \sigma_z^i\right)^2,
\end{equation}
where $\delta_k$ is the detuning to the mode $k$, $\eta_k$ the Lamb-Dicke parameter, $\Omega_i$ the $411\,$nm-laser-induced AC Stark shift on the ion $i$, and $b_{ik}$ the normalized mode vector. Therefore when $\delta_k>0$, the highest eigenstate of $H_0$, or the ground state of $-H_0$, can be simply expressed as $\sigma_z^i=\pm\mathrm{sign}(b_{ik})$, thus imprinting the phonon mode structure into the observed spatial correlation.

Combining this classically solvable ground state and the spatial resolution, we can verify the quantum simulation result by comparing with the calculated phonon mode. Beyond the relatively simple mode patterns in Fig.~\ref{fig2}\textbf{a}-\textbf{d}, we can also couple to a phonon mode with spatial structures both along the major and the minor axes of the ion crystal in Fig.~\ref{fig2}\textbf{e}-\textbf{h}. The typical single-shot measurement result (Fig.~\ref{fig2}\textbf{g}) agrees well with the theoretically calculated mode vector (Fig.~\ref{fig2}\textbf{h}). We further show the average spin-spin correlation from 100 repetitions in Fig.~\ref{fig2}\textbf{e}. Because of the 2D spatial pattern, the correlation appears to be noisy in Fig.~\ref{fig2}\textbf{e} as we flatten the 2D structure into 1D in the rows and columns of the matrix. However, if we simply rearrange the indices of the ions into the descending order of $b_{ik}$, we obtain the matrix in Fig.~\ref{fig2}\textbf{f}. Here we have two groups of ions with positive and negative $b_{ik}$'s, which explains the positive correlation within each group, and the negative correlation between them. Again, this verifies the agreement between the theoretical and the experimental results. Note that here when computing the phonon mode structure, we simply assume a harmonic trap and use the measured ion positions as the starting point to search the theoretical equilibrium configuration. This can be improved by fitting the anharmonicity over the large crystal using the measured transverse mode frequencies.

\begin{figure*}[!tbp]
	\centering
	\includegraphics[width=0.9\linewidth]{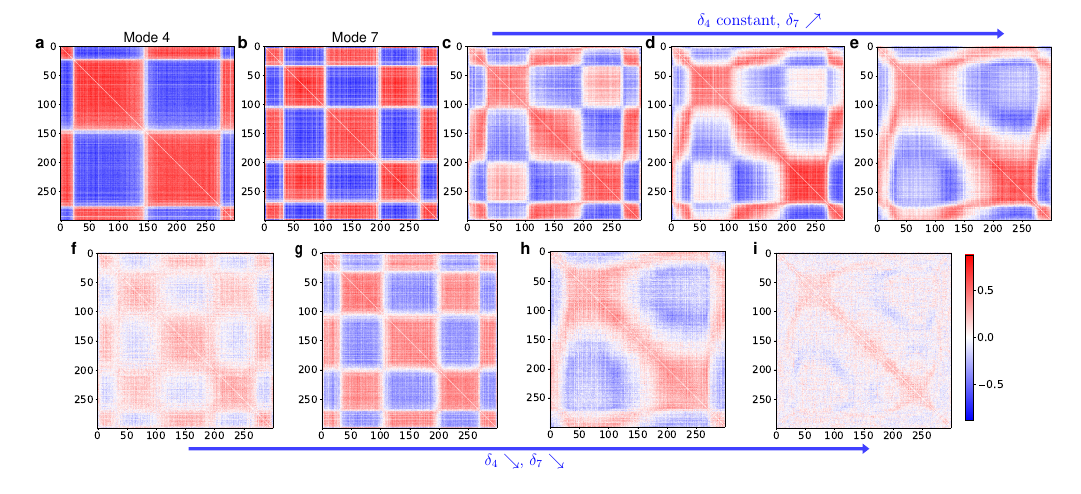}
	\caption {\textbf{Quantum simulation of frustrated Ising model.} \textbf{a}, \textbf{b}, The covariance matrices for the same $N=300$ crystal as Fig.~\ref{fig2} when the beat note of the $411\,$nm laser is $2\pi\times 1\,$kHz above the 4th highest phonon mode and $2\pi\times 1\,$kHz above the 7th highest phonon mode, respectively. \textbf{c}-\textbf{e}, Two beat note frequencies of the $411\,$nm laser are applied to couple both phonon modes simultaneously. From left to right, we fix the detuning to the 4th mode as $\delta_4=2\pi\times 0.75\,$kHz, and increase the detuning to the 7th mode as $\delta_7=2\pi\times [0.50,\,0.75,\,1.00]\,$kHz. On the two ends, the ground states are dominated by the two modes, respectively, while in the middle there is frustration due to the competition between the two configurations.
\textbf{f}-\textbf{i}, Again we couple both phonon modes simultaneously. From left to right, we decrease the detunings to the two phonon sidebands simultaneously as $(\delta_4,\delta_7)/2\pi=(2.5, 1.5), (1.5, 0.5), (0.5,-0.5), (-0.5, -1.5) \,$kHz. When $\delta_4,\delta_7>0$ (\textbf{g}), the ground state is dominated by the mode $7$ under the chosen parameters, similar to \textbf{c}. When detunings further increase (\textbf{f}), the Ising coupling decreases, and so do the measured correlations due to the stronger nonadiabatic excitation. When $\delta_4>0$ and $\delta_7<0$ (\textbf{h}), the ground state is dominated by the mode $4$, but also feels the frustrated Ising coupling from the mode $7$ under negative detuning. Finally, when $\delta_4,\delta_7<0$ (\textbf{i}), both contributions to the Ising coupling coefficients are strongly frustrated and it is difficult to theoretically predict the result. All the correlations are averaged over 100 samples.
}\label{fig4}
\end{figure*}

In all the above cases we couple dominantly to a single phonon mode with positive detuning, so that the Ising coupling shows no frustration and the ground state can be easily understood from the phonon mode structure. We further engineer frustrated Ising models in Fig.~\ref{fig4} by coupling to more phonon modes simultaneously, or by setting a negative detuning. In Fig.~\ref{fig4}\textbf{a} and \textbf{b} we show the spin-spin correlation when coupling to the mode $4$ and the mode $7$ individually.
Then we couple to the two modes simultaneously via the two pairs of frequency components in the $411\,$nm laser in Fig.~\ref{fig1}\textbf{b}. From Fig.~\ref{fig4}\textbf{c} to \textbf{e}, we fix the detuning $\delta_4$ to the mode $4$ and increase the detuning $\delta_7$ to the mode $7$, such that their relative contribution to the Ising Hamiltonian changes. For small $\delta_7>0$ (Fig.~\ref{fig4}\textbf{c}), the mode $7$ dominates and the correlation pattern is similar to Fig.~\ref{fig4}\textbf{b}; for large $\delta_7$ (Fig.~\ref{fig4}\textbf{e}), the mode $4$ dominates and the correlation pattern resembles Fig.~\ref{fig4}\textbf{a}; while in the middle with roughly equal contribution from both modes, we see competition from the two patterns and the correlation becomes their mixture. From Fig.~\ref{fig4}\textbf{f} to \textbf{i}, we decrease both $\delta_4$ and $\delta_7$ from positive to negative. In Fig.~\ref{fig4}\textbf{g} with $\delta_4=2\pi\times 1.5\,$kHz and $\delta_7=2\pi\times 0.5\,$kHz, the situation is similar to Fig.~\ref{fig4}\textbf{c} dominated by the mode $7$. If we further increase $\delta_4$ and $\delta_7$ to Fig.~\ref{fig4}\textbf{f}, the generated Ising coupling decreases as well as the energy gap, making it more difficult to prepare the ground state adiabatically and thus the correlation also shrinks. On the other hand, if we decrease the detuning to $\delta_4=2\pi\times 0.5\,$kHz and $\delta_7=-2\pi\times 0.5\,$kHz (Fig.~\ref{fig4}\textbf{h}), the ground state becomes dominated by the mode $4$ with positive detuning, but is also influenced by the weak frustrated Ising coupling from the negatively detuned mode $7$.
Finally, when both $\delta_4$ and $\delta_7$ are negative (Fig.~\ref{fig4}\textbf{i}), the Ising model becomes highly frustrated and the correlation in the slowly ramped state also becomes much weaker, with no simple analytical explanation for the observed pattern. In Supplementary Information we further compare these results with numerical results by the classical simulated annealing algorithm to show that our quantum algorithm gives the correct pattern for the ground state in the ``simple'' case, while its output is difficult to predict in the ``hard'' case with strong frustration.

\begin{figure}[!tbp]
	\centering
	\includegraphics[width=\linewidth]{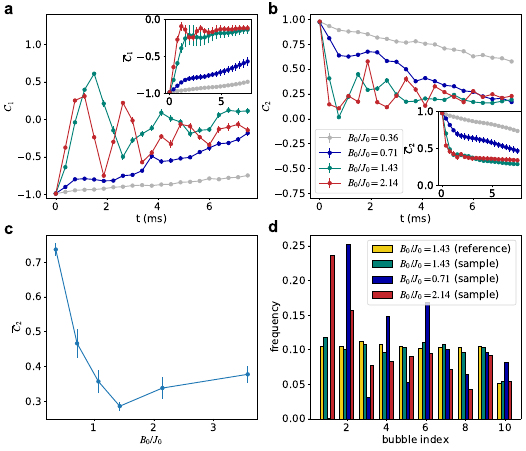}
	\caption{\textbf{Quench dynamics and quantum sampling.} \textbf{a}, Average single-spin dynamics $C_1(t) \equiv \sum_i \langle \sigma_z^i(t)\rangle / N$ under various values of the transverse field. Error bar shows statistical fluctuation from 100 shots. Here we fix the Ising coupling by tuning the laser beatnote $2\pi\times 4\,$kHz above the COM mode and obtain a Kac normalized $J_0=2\pi\times 0.31\,$kHz. Inset shows the cumulative time-averaged value $\bar{C}_1(t)\equiv \int_0^t C_1(\tau) d\tau / t$, which is more robust against slow parameter drift. Error bars represent one standard error from three groups of data taken at different time. \textbf{b}, Similar plot for average two-spin dynamics $C_2\equiv \sum_{ij} \langle \sigma_z^i \sigma_z^j\rangle / N^2$. \textbf{c}, $\bar{C}_2(T)$ vs. transverse field $B_0$ for a fixed evolution time $T=7.5\,$ms. A dip can be observed near $B_0/J_0=1.5$, indicating a dynamical phase transition.
\textbf{d}, Histogram of the sampled $300$-qubit data in 10 coarse-grained ``bubbles''. The data are sampled from three different quantum states as the time evolution of $T=6\,$ms under three different transverse-fields $B_0/J_0=0.71,\,1.43,\,2.14$. We first collect about $5000$ samples at $B_0/J_0=1.43$ as the reference to design the bubbles for coarse graining and to estimate the probability distribution in these bubbles. Then we collect $1000$ samples from $B_0/J_0=0.71,\,1.43,\,2.14$, respectively, and compare their distributions with the reference. The samples from $B_0/J_0=1.43$ agree with the reference at a $p$-value of 0.8, while the samples from $B_0/J_0=0.71$ and $2.14$ are rejected at a $p$-value below $10^{-49}$.
}\label{fig5}
\end{figure}

Another task that is enabled by our large qubit capacity and individual single-shot readout, and is challenging for direct simulation by classical computers, is to sample from the quantum states after many-body Hamiltonian evolution, which belongs to the quantum sampling problem \cite{Smith2019,PRXQuantum.4.020310}. Specifically, we consider the quench dynamics of a 2D transverse-field Ising model with long-range interaction by turning on the Ising coupling $J_{ij}$ and the transverse field $B_0$ simultaneously. By choosing the initialization and readout bases along the $\sigma_z$ direction, our model is similar to that in Ref.~\cite{zhang2017Observation} with the $\sigma_x$ and $\sigma_z$ bases exchanged. Actually, from the dynamics of single-spin and two-spin observables in Fig.~\ref{fig5}\textbf{a}-\textbf{c}, similar signals for a dynamical phase transition can be observed: At small $B_0$, the spins remain pinned near their initially polarized direction, and we get nonzero single-spin $C_1\equiv \sum_i \langle \sigma_z^i\rangle / N$ and two-spin $C_2\equiv \sum_{ij} \langle \sigma_z^i \sigma_z^j\rangle / N^2$ expectations after long evolution time; At large $B_0$, spins precess around the transverse field and the $Z_2$ symmetry of the Ising Hamiltonian is recovered, which gives time-averaged $C_1\to 0$ and $C_2\to 1/2$ as the spins rotate collectively between $\pm 1$ \cite{zhang2017Observation}. In the experiment, we get deviation from the above theoretical values for large $B_0$, which can be explained by a small uncompensated longitudinal field which breaks the $Z_2$ symmetry, and by the nonuniformity of the transverse field over the large ion crystal causing different precession speed for different spins.

While the above single-spin and two-spin dynamics can be understood semi-quantitatively and provide a way to calibrate our quantum simulator, in general it is much more difficult to directly simulate the many-body quantum dynamics by a classical computer and to sample from the probability distribution of the multi-qubit measurement outcome. On the other hand, it is straightforward to obtain such samples from a quantum simulator. In Fig.~\ref{fig5}\textbf{d} we generate such samples for three different values of the transverse field $B_0$, and we perform a coarse-grained analysis \cite{wang2016coarse} (see Supplementary Information for more details) to show that their underlying probability distributions are distinct, thus are nontrivial and are not dominated by experimental decoherence.

\section{Discussion and outlook}
In this work, we achieve the stable trapping of a 2D crystal of above $500$ ions, and demonstrate the quantum simulation of $300$ ions with individual state detection. Currently this smaller ion number is chosen according to our available $411\,$nm laser power to cover the whole ion crystal for strong Ising coupling, and is not a fundamental limit. To further scale up the system to thousands of ions, we may also perform sympathetic cooling on a few ions with optimized locations \cite{PhysRevLett.127.143201} to maintain its stability, while using the dual-type qubit scheme to avoid crosstalk errors on the ions carrying quantum information as we have demonstrated recently in small systems \cite{yang2022realizing}.

We create frustrated Ising model Hamiltonian by coupling to up to two phonon modes. By adding more frequency components into the $411\,$nm laser \cite{Korenblit_2012,wu2023qubits} or by applying a spatial gradient of the AC Stark shift \cite{PhysRevX.13.021021}, it will be possible to engineer more complicated coupling coefficients, and thus to simulate rich quantum dynamics that are intractable for classical computers \cite{RevModPhys.86.153} and to execute NISQ algorithms \cite{RevModPhys.94.015004} like quantum annealing \cite{Hauke_2020} and variational quantum optimization \cite{cerezo2021variational}. Furthermore, in the future by integrating the 2D laser addressing into the system \cite{pu2017experimental,PhysRevLett.125.150505}, our 2D ion crystal may also support high-fidelity two-qubit entangling gates mediated by the transverse phonon modes \cite{wang2015quantum,PhysRevA.103.022419}, thus makes a promising way to extend the scale of ion trap quantum computers.

\bigskip

\textbf{Data Availability:}
The source data supporting this work are available from the figshare data repository (https://figshare.com) under https://doi.org/10.6084/m9.figshare.25572603.

\textbf{Code Availability:} The codes supporting this work are available from the corresponding author upon request.

\textbf{Acknowledgements:} This work was supported by Innovation Program for Quantum Science and Technology (2021ZD0301601,2021ZD0301605), Tsinghua University Initiative Scientific Research Program, and the Ministry of Education of China. L.M.D. acknowledges in addition support from the New Cornerstone Science Foundation through the New Cornerstone Investigator Program. Y.K.W. acknowledges in addition support from Tsinghua University Dushi program and the start-up fund.

\textbf{Competing interests:} W.Q.L., R.Y., Y.W., B.W.L. and W.X.G. are affiliated with HYQ Co. Y.K.W., W.Q.L., R.Y., Y.W., B.W.L., Y.Z.X., W.X.G., B.X.Q., Z.C.Z., L.H. and L.M.D. hold shares with HYQ Co. The other authors declare no competing interests.

\textbf{Author Information:} Correspondence and requests for materials should be addressed to L.M.D.
(lmduan@tsinghua.edu.cn).

\textbf{Author Contributions:} L.M.D. proposed and supervised the project. S.A.G., J.Y., L.Z., W.Q.L., R.Y., Y.W., R.Y.Y., Y.J.Y., Y.L.X., B.W.L., Y.H.H., Y.Z.X., W.X.G., C.Z., B.X.Q., Z.C.Z., L.H. carried out the experiment. S.A.G., Y.K.W., J.Y. analyzed the data and did the associated theory. Y.K.W., S.A.G., and L.M.D. wrote the manuscript.


\end{document}


\title{Supplementary Information for\\A Site-Resolved 2D Quantum Simulator with Hundreds of Trapped Ions\\
}

\author{S.-A. Guo}
\affiliation{Center for Quantum Information, Institute for Interdisciplinary Information Sciences, Tsinghua University, Beijing 100084, PR China}

\author{Y.-K. Wu}
\affiliation{Center for Quantum Information, Institute for Interdisciplinary Information Sciences, Tsinghua University, Beijing 100084, PR China}
\affiliation{Hefei National Laboratory, Hefei 230088, PR China}

\author{J. Ye}
\affiliation{Center for Quantum Information, Institute for Interdisciplinary Information Sciences, Tsinghua University, Beijing 100084, PR China}

\author{L. Zhang}
\affiliation{Center for Quantum Information, Institute for Interdisciplinary Information Sciences, Tsinghua University, Beijing 100084, PR China}

\author{W.-Q. Lian}
\affiliation{HYQ Co., Ltd., Beijing 100176, PR China}

\author{R. Yao}
\affiliation{HYQ Co., Ltd., Beijing 100176, PR China}

\author{Y. Wang}
\affiliation{Center for Quantum Information, Institute for Interdisciplinary Information Sciences, Tsinghua University, Beijing 100084, PR China}
\affiliation{HYQ Co., Ltd., Beijing 100176, PR China}

\author{R.-Y. Yan}
\affiliation{Center for Quantum Information, Institute for Interdisciplinary Information Sciences, Tsinghua University, Beijing 100084, PR China}

\author{Y.-J. Yi}
\affiliation{Center for Quantum Information, Institute for Interdisciplinary Information Sciences, Tsinghua University, Beijing 100084, PR China}

\author{Y.-L. Xu}
\affiliation{Center for Quantum Information, Institute for Interdisciplinary Information Sciences, Tsinghua University, Beijing 100084, PR China}

\author{B.-W. Li}
\affiliation{HYQ Co., Ltd., Beijing 100176, PR China}

\author{Y.-H. Hou}
\affiliation{Center for Quantum Information, Institute for Interdisciplinary Information Sciences, Tsinghua University, Beijing 100084, PR China}

\author{Y.-Z. Xu}
\affiliation{Center for Quantum Information, Institute for Interdisciplinary Information Sciences, Tsinghua University, Beijing 100084, PR China}

\author{W.-X. Guo}
\affiliation{HYQ Co., Ltd., Beijing 100176, PR China}

\author{C. Zhang}
\affiliation{Center for Quantum Information, Institute for Interdisciplinary Information Sciences, Tsinghua University, Beijing 100084, PR China}

\author{B.-X. Qi}
\affiliation{Center for Quantum Information, Institute for Interdisciplinary Information Sciences, Tsinghua University, Beijing 100084, PR China}

\author{Z.-C. Zhou}
\affiliation{Center for Quantum Information, Institute for Interdisciplinary Information Sciences, Tsinghua University, Beijing 100084, PR China}
\affiliation{Hefei National Laboratory, Hefei 230088, PR China}

\author{L. He}
\affiliation{Center for Quantum Information, Institute for Interdisciplinary Information Sciences, Tsinghua University, Beijing 100084, PR China}
\affiliation{Hefei National Laboratory, Hefei 230088, PR China}

\author{L.-M. Duan}
\email{lmduan@tsinghua.edu.cn}
\affiliation{Center for Quantum Information, Institute for Interdisciplinary Information Sciences, Tsinghua University, Beijing 100084, PR China}
\affiliation{Hefei National Laboratory, Hefei 230088, PR China}
\affiliation{New Cornerstone Science Laboratory, Beijing 100084, PR China}

\maketitle

\makeatletter
\renewcommand{\thefigure}{S\arabic{figure}}
\renewcommand{\thetable}{S\arabic{table}}
\renewcommand{\theHfigure}{S.\arabic{figure}}
\renewcommand{\theHtable}{S.\arabic{table}}
\renewcommand{\fnum@figure}{Fig.~\thefigure}
\renewcommand{\fnum@table}{Table.~\thetable}
\makeatother


\section{Experimental setup}
\subsection{Monolithic 3D ion trap}
We use a monolithic 3D Paul trap \cite{Brownnutt_2006,https://doi.org/10.1002/qute.202000068} (see Fig.~1\textbf{a} of main text) with an RF frequency of $\omega_{\mathrm{rf}}=2\pi\times 35.280\,$MHz to hold 2D crystals of ${}^{171}\mathrm{Yb}^+$ ions. The trap is fabricated on a $0.508\times23\times23\,$mm$^3$ alumina chip by laser cutting. Gaps between electrodes are of $50\,\mu$m width and about $10\,\mu$m fabrication accuracy, and the segments of the DC electrodes have a width of $600\,\mu$m. The distance between the trap center and the RF electrode is $150\,\mu$m. The monolithic fabrication can largely eliminate the assembly error compared with our previous blade traps, and can have stronger and more symmetric confinement compared with our previous surface traps. The voltage on the $4\times7=28$ DC segments can be controlled independently, together with an overall DC bias on the RF electrodes. These provide us with sufficient degrees of freedom to control the shape of the 2D crystal and to eliminate its micromotion perpendicular to the plane. To suppress the collision rate of the ion crystal with the background gas molecules and thus to improve its stability, we place the trap at a cryogenic temperature of $6.1\,$K. More details about the stability of ion crystals can be found in Sec.~\ref{sec:stability}.

\subsection{Laser configuration}
\begin{figure}[htbp]
   \includegraphics[width=0.6\linewidth]{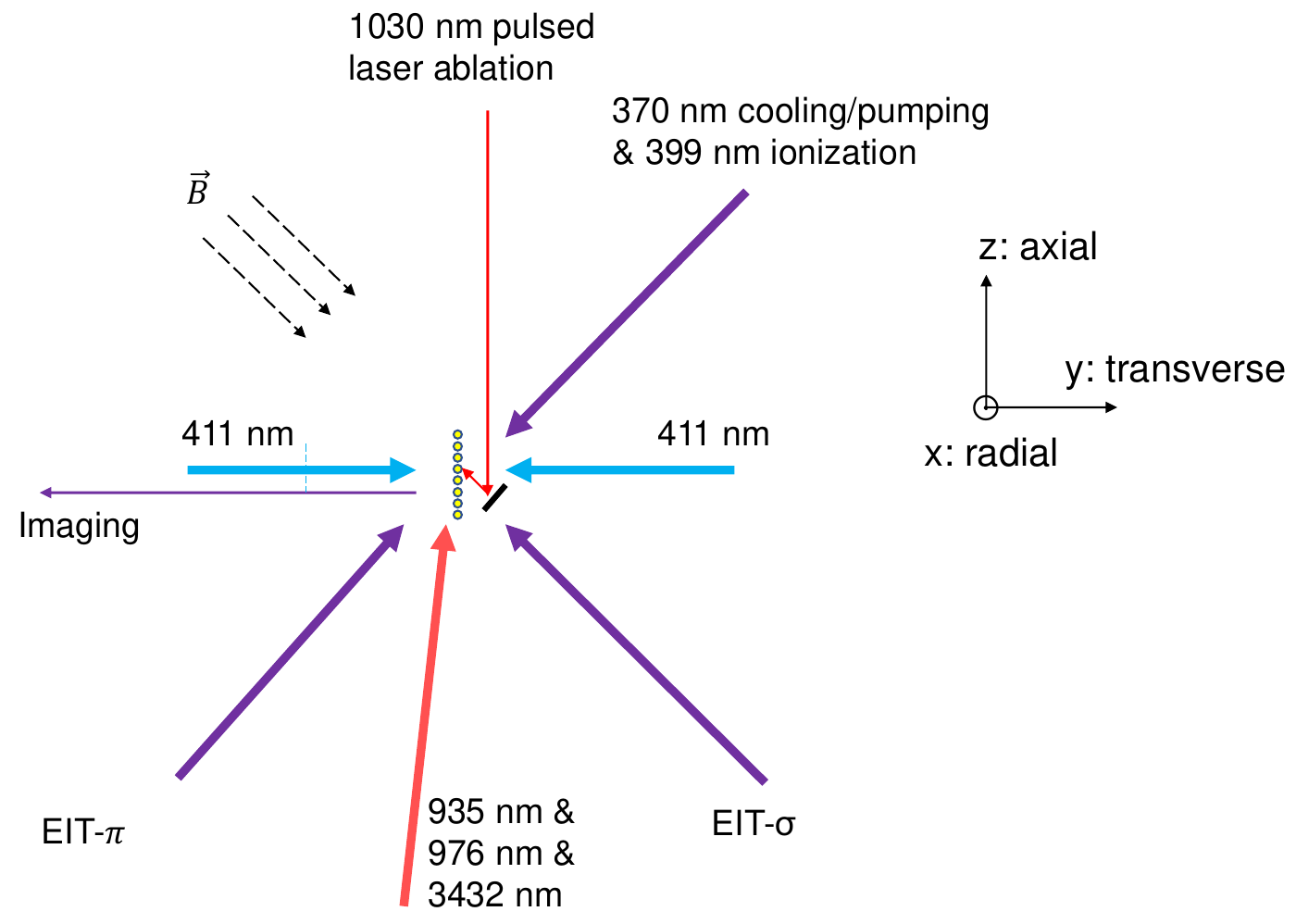}
   \caption{Laser configuration.
   \label{figS1}}
\end{figure}

All the laser beams propagate in the micromotion-free directions in the $y$-$z$ plane, as shown in Fig.~\ref{figS1}. This makes their effect insensitive to the inevitable micromotion of the 2D crystal along the $x$ direction.

To load ions into the trap, first we generate neutral atomic beams aiming at the trap center by laser ablation on a target of isotope-enriched $^{171}\mathrm{Yb}$ metal using $1030\,$nm pulsed laser. The $399\,$nm ionization laser is perpendicular to the atomic beam, which eliminates the Doppler shift and thus enables better isotope selectivity.
The $935\,$nm, $976\,$nm and $3432\,$nm laser beams are combined together and enter the trap at a small angle with respect to the $z$-axis to cover the whole ion crystal. They are used to repump the population on $^{2}D_{3/2}$, $^{2}D_{5/2}$ and $^{2}F_{7/2}$, respectively, back to $^{2}S_{1/2}$.
The $370\,$nm laser has its main component for Doppler cooling, and a largely red-detuned component as the protection beam. This cooling beam can directly cool down the phonon modes in the micromotion-free $y$ and $z$ directions. Since the in-plane motion of the ions in the $x$ and $z$ directions are coupled together, we will also be able to cool down most of the $x$ modes. An exception is the center-of-mass (COM) mode in the $x$ direction, which is not coupled to the motion in the $y$ direction in the ideal case. Nevertheless, in the experiment we observe that this mode can still be cooled, which may come from the anharmonicity of the trap or a small angle of the cooling laser away from the $y$-$z$ plane. Furthermore, by turning on the $14.7\,$GHz and $2.1\,$GHz electro-optic modulators (EOMs), the $370\,$nm beam can also be switched for optical pumping and qubit state detection.
Two additional $370\,$nm laser beams, perpendicular to each other with $\pi$ and $\sigma^+$ polarizations, respectively, are used for EIT cooling \cite{PhysRevLett.125.053001}. They have a blue detuning of about $86\,$MHz from the transition between $|^{2}S_{1/2},F=1,m_F=0\rangle$ ($|^{2}S_{1/2},F=1,m_F=-1\rangle$) and $|^{2}P_{1/2},F=0,m_F=0\rangle$.

A pair of counter-propagating $411\,$nm laser beams are applied perpendicular to the 2D crystal, with a linewidth of about $1\,$kHz. Either of the two beams can be used for sideband cooling and electron shelving to the $D_{5/2}$ levels. Together, they can generate a spin-dependent force which can further lead to the Ising coupling as we describe in Sec.~\ref{sec:411}. The waist diameters of the two counter-propagating beams are about $50\times 330\,\mu$m$^2$ and $42\times 380\,\mu$m$^2$ respectively, which can cover the whole 2D ion crystal. Both beams are polarized in the $z$ direction. Together with a magnetic field of $4.6\,$G at an angle of $45^\circ$ to the $y$ and $z$ directions, this geometry can maximize the transition matrix element between the $|^{2}S_{1/2},F=0,m_F=0\rangle$ and $|^{2}D_{5/2},F=2,m_F=0\rangle$ states.

\subsection{Imaging system}
The imaging system is aligned perpendicular to the ion crystal with an NA of 0.33. We use a CMOS camera to collect the site-resolved fluorescence from individual ions. Each image has a size of $552\times 88$ pixels, and the typical distance between adjacent ions is about $7.7\,$pixels, or about $4.1\,\mu$m given our magnification of $0.535\,\mu$m/pixel. To obtain the image for the $512$-ion crystal in Fig.~1\textbf{c} of the main text, we use an exposure time of $150\,$ms. Later to perform the single-shot state detection, we first shelve the $|^{2}S_{1/2},F=0,m_F=0\rangle$ state to the $D_{5/2}$ and $F_{7/2}$ levels by the global $411\,$nm and $3432\,$nm laser \cite{edmunds2020scalable,yang2022realizing}, and then detect the fluorescence of the ions under the global $370\,$nm laser with an exposure time of $1.5\,$ms. The state detection infidelity is about $1\%$ for the dark state due to the imperfect shelving under inhomogeneous laser beams, and about $0.1\%$ for the bright state.

\subsection{Trap parameters}
We encode the qubit state in the $|0\rangle\equiv|^{2}S_{1/2},F=0,m_F=0\rangle$ and $|1\rangle\equiv|^{2}S_{1/2},F=1,m_F=0\rangle$ levels of the ${}^{171}\mathrm{Yb}^+$ ions. The single-ion coherence time is measured to be $T_2\approx 0.5\,$s. The motional coherence time for the transverse mode is directly measured to be about $1.5\,$ms, and can be extended to $10\,$ms with a $50\,$Hz line-trigger and a spin echo.

\begin{figure}[htbp]
   \includegraphics[width=\linewidth]{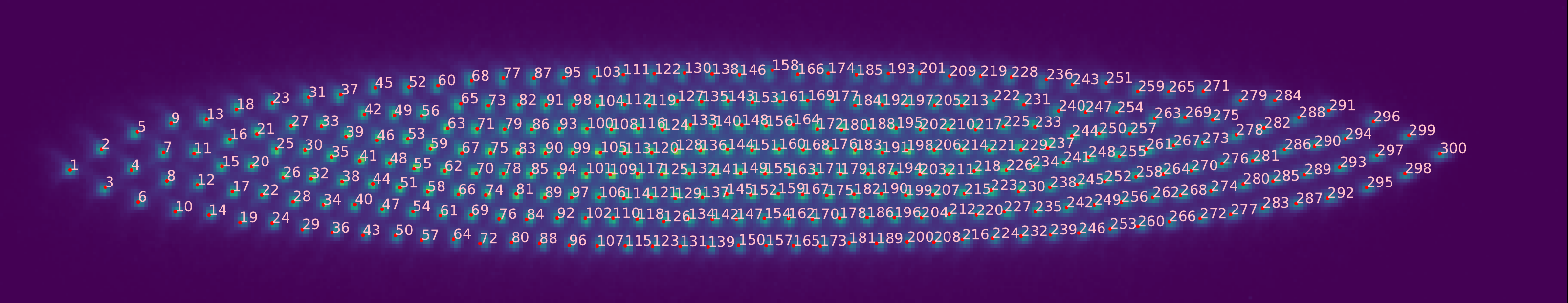}
   \caption{The experimental image of the 2D crystal of $N=300$ ions and their labels according to the $z$ coordinates.
   \label{figS2}}
\end{figure}

To hold the 2D crystal of $N=512$ ions, we use a trap frequency of $(\omega_x,\omega_y,\omega_z)=2\pi\times(0.60,2.164,0.144)\,$MHz, and when holding the crystal of $N=300$ ions, we slightly increase the in-plane trapping and decrease the transverse trapping to $(\omega_x,\omega_y,\omega_z)=2\pi\times(0.69,2.140,0.167)\,$MHz. The 2D ion crystal can be held for days under global Doppler cooling, and its dark lifetime with the cooling laser turned off is on the timescale of seconds. The label for each ion in the $N=300$ ion crystal in the main text is shown in Fig.~\ref{figS2}, which is in ascending order of their $z$ coordinates. Note that the above transverse mode frequency $\omega_y$ can be measured from the sideband of the $411\,$nm laser, while the in-plane trap frequencies $\omega_x$ and $\omega_z$ are obtained from parametric resonance using a single ion. Since these in-plane trap frequencies are subjected to spatial inhomogeneity over the large 2D crystal, we further adjust their values slightly to match the shape of the theoretically computed equilibrium configuration of the 2D crystal with the measured image as much as possible. The phonon mode frequencies computed in this way typically agree with the measured values with a deviation below $1$-$2\,$kHz in the high-frequency range, say, within $150\,$kHz below the COM mode. On the other hand, the low-frequency modes are more sensitive to the local properties of the trap and we observe larger deviation between the theoretical and the experimental results.

\subsection{Estimating phonon number}
We follow the method of Ref.~\cite{PhysRevLett.129.140501} to estimate the average phonon number in each mode, with small modification because here we are using $411\,$nm laser to couple the electric quadrupole transition between $|\downarrow\rangle\equiv|^{2}S_{1/2},F=0,m_F=0\rangle$ and $|\uparrow\rangle\equiv|^{2}D_{5/2},F=2,m_F=0\rangle$, rather than the Raman transition of $355\,$nm laser between $|0\rangle\equiv|^{2}S_{1/2},F=0,m_F=0\rangle$ and $|1\rangle\equiv|^{2}S_{1/2},F=1,m_F=0\rangle$. Specifically, we initialize all the ions in $|\downarrow\rangle^{\otimes N}$, apply a weak global $411\,$nm laser beam with a short duration around the red and the blue motional sidebands, and measure the final spin states by the fluorescence under a global $370\,$nm laser.
As shown in Fig.~1\textbf{d} and \textbf{e} in the main text, under weak driving, different phonon sidebands can be well distinguished, so the effects of the off-resonant terms can be neglected.

For completeness, below we summarize the derivations in the Supplementary Information of Ref.~\cite{PhysRevLett.129.140501}.
Consider a driving laser resonant to the red motional sideband of the $k$-th mode. We get the Hamiltonian
\begin{equation}
H_r = \sum_i \eta_k b_{ik} \Omega_i (a_k \sigma_i^+ + a_k^\dag \sigma_i^-),
\end{equation}
where $\eta_k$ is the Lamb-Dicke parameter of mode $k$, $\Omega_i$ the carrier Rabi frequency of the driving laser on ion $i$, $b_{ik}$ the mode vector, $a_k$ ($a_k^\dag$) the annihilation (creation) operator of mode $k$, and $\sigma_i^+\equiv|\uparrow\rangle_i\langle\downarrow|$ ($\sigma_i^-\equiv|\downarrow\rangle_i\langle\uparrow|$) the upper (lower) operator of ion $i$.

Suppose initially there are $n$ phonons in the $k$-th mode. When applying this Hamiltonian for a short time $T$, the evolution of the system can be approximated as
\begin{equation}
|\Psi^r(T)\rangle \approx  |\downarrow\rangle^{\otimes N}|n\rangle -i \sqrt{n} T \sum_i \eta_k b_{ik} \Omega_i |\downarrow\rangle^{\otimes (i-1)}|\uparrow\rangle |\downarrow\rangle^{\otimes (N-i)} |n-1\rangle,
\end{equation}
where we have neglected the weaker multi-spin excitations. Averaging over the phonon number distribution $\{p_n\}$, we get the probability for a single-spin excitation as
\begin{equation}
P_r(T) \approx  \sum_n p_n \sum_i \left(\sqrt{n} T \eta_k b_{ik} \Omega_i\right)^2 = \bar{n} \sum_i \left(T \eta_k b_{ik} \Omega_i\right)^2.
\end{equation}

Similarly, when the same driving laser is applied on the blue sideband, we have
\begin{equation}
H_b = \sum_i \eta_k b_{ik} \Omega_i (a_k^\dag \sigma_i^+ + a_k \sigma_i^-),
\end{equation}
and
\begin{equation}
P_b(T) \approx  \sum_n p_n \sum_i \left(\sqrt{n+1} T \eta_k b_{ik} \Omega_i\right)^2 = (\bar{n}+1) \sum_i \left(T \eta_k b_{ik} \Omega_i\right)^2.
\end{equation}
Therefore we can obtain $\bar{n}$ from $P_r(T)$ and $P_b(T)$ as $\bar{n}=P_r(T)/[P_b(T)-P_r(T)]$. In particular, for a low average phonon number $\bar{n}\ll 1$, we have $\bar{n}\approx P_r(T)/P_b(T)$. In the above derivation, we assume resonant driving to the red and the blue sidebands, but the same ratio can be obtained if we consider the same detuning to the two sidebands as long as the excitation is still dominated by a single phonon mode. This will allow us to average over a few data points around each peak to suppress the statistical fluctuation. Also note that the above derivations require $(\bar{n}+1) \sum_i \left(T \eta_k b_{ik} \Omega_i\right)^2 \ll 1$. For example, for the Doppler cooling results in Fig.~1\textbf{d} and \textbf{e} in the main text with a high phonon number, we typically obtain higher blue sidebands than the red sidebands due to the non-negligible multi-spin excitations. Finally, as shown in Supplementary Information of Ref.~\cite{PhysRevLett.129.140501} through numerical examples, the validity of this approximation can be extended to about $\eta_k\bar{\Omega}T\lesssim \pi/2$ to estimate the order of magnitude, so as to allow stronger red sideband signal and to increase the signal-to-noise ratio.

For experimental convenience, we estimate the ratio between $P_r(T)$ and $P_b(T)$ from the overall photon counts under the $370\,$nm laser rather than distinguishing individual spin states. Suppose the average photon number from ion $i$ in the bright state $|\downarrow\rangle$ is $N_i$, and almost zero when shelved to the $|\uparrow\rangle$ state. Also, suppose there are $N_0$ dark counts for the CMOS camera. Then we can estimate the average photon counts under red sideband and blue sideband driving as
\begin{equation}
N_r = N_0 +  \sum_i \left[1 - \bar{n}\left(T \eta_k b_{ik} \Omega_i\right)^2\right] N_i,
\end{equation}
and
\begin{equation}
N_b = N_0 + \sum_i \left[1 - (\bar{n}+1) \left(T \eta_k b_{ik} \Omega_i\right)^2\right] N_i.
\end{equation}

In the experiment, we first measure the maximal photon counts $N_{\mathrm{max}}=N_0+\sum_i N_i$ when all the ions are initialized in $|\downarrow\rangle$. Then we have
\begin{equation}
N_{\mathrm{max}} - N_r = \bar{n} \sum_i \left(T \eta_k b_{ik} \Omega_i\right)^2 N_i,
\end{equation}
and
\begin{equation}
N_{\mathrm{max}} - N_b = (\bar{n}+1) \sum_i \left(T \eta_k b_{ik} \Omega_i\right)^2 N_i,
\end{equation}
which allow us to estimate the average phonon number $\bar{n}$ in a similar way as using $P_r(T)$ and $P_b(T)$. Note that using this method, most of the population will be in the bright state and the weak spin excitation only leads to small decay in the photon counts. Therefore, the result can be sensitive to the slow drift in the laser intensity and the statistical fluctuation of the photon number following a Poisson distribution. Therefore, we choose to measure $N_{\mathrm{max}}$ before each data point individually to compensate the drift of the photon scattering rate, and we repeat the measurement for 1000 times to suppress the statistical fluctuation.

\begin{figure}[htbp]
   \includegraphics[width=0.8\linewidth]{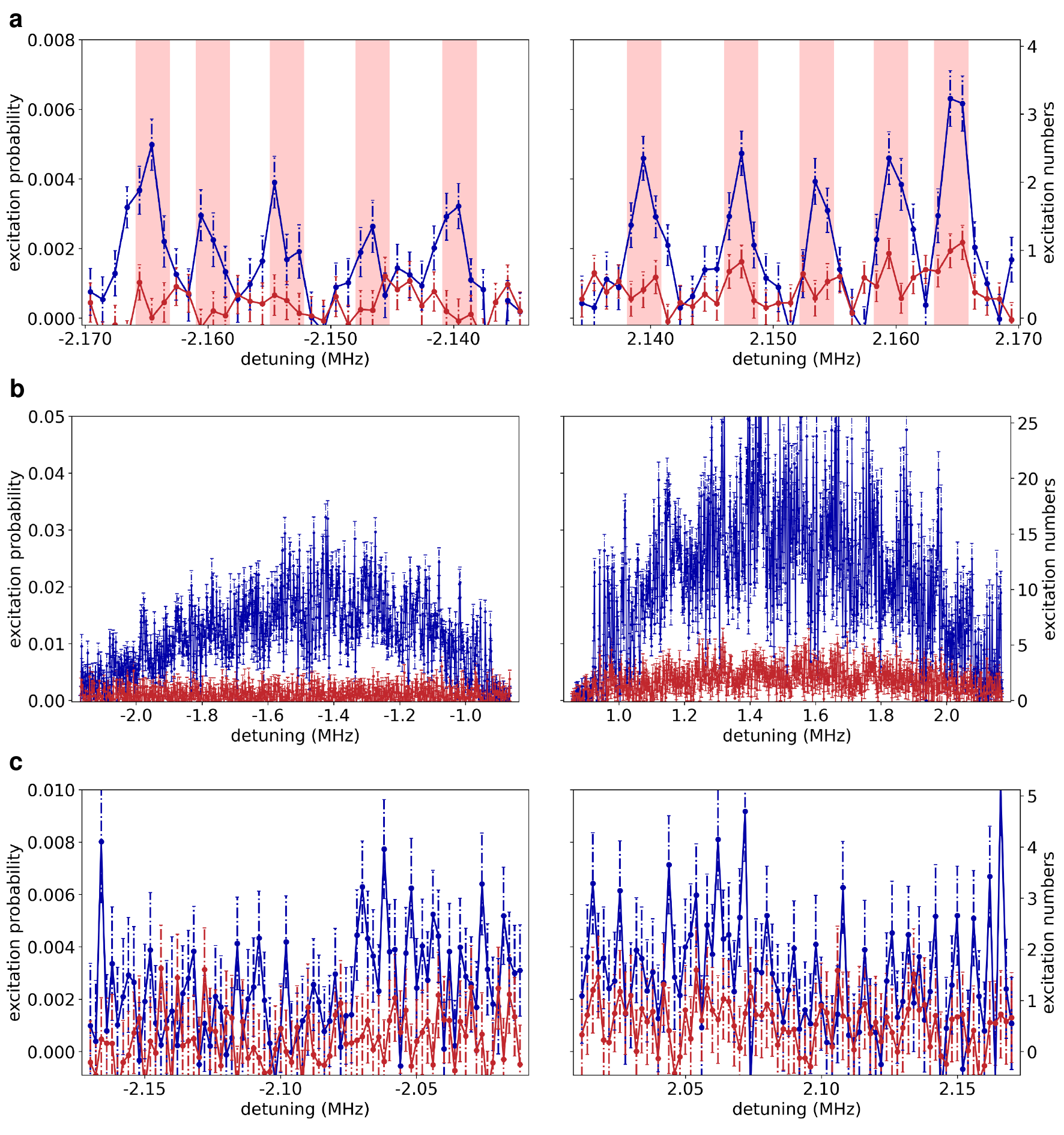}
   \caption{\textbf{a}, The same plot as Fig.~1\textbf{d} and \textbf{e} of the main text, with the three data points around each mode shaded for estimating the phonon number. \textbf{b}, Spectra for the red and blue motional sidebands of all the transverse modes for $N=512$ ions. \textbf{c}, The highest $2\pi\times 150\,$kHz below the COM mode subjected to sideband cooling. The blue (red) points are the spectra after Doppler cooling (EIT and sideband cooling).
   \label{figS3}}
\end{figure}

Following this method, we estimate the average phonon number for the five modes shown in Fig.~1\textbf{d} and \textbf{e} of the main text to be $\bar{n}_k=(0.4 \pm 0.3, 0.0 \pm 0.3, 0.8\pm 1.2, 0.9 \pm 1.1, 0.1\pm0.5)$ from the COM mode to the lower modes, where we average over 3 data points around each phonon mode (see Fig.~\ref{figS3}\textbf{a}) to suppress the statistical error and the small drift in the mode frequencies. Note that here we have negative points in the red sideband after subtracting the background due to such random fluctuation. Also note that the shape of these spectra after sideband cooling suggests that there is still non-negligible statistical fluctuation, which will prevent us from obtaining a red sideband below the error bar. Therefore, the actual phonon number in these modes can be lower than the measured values. In Fig.~\ref{figS3}\textbf{b} and \textbf{c} we further show the spectra for all the $N=512$ transverse modes and a zoom-in for the highest $2\pi\times 150\,$kHz range for which we perform the sideband cooling. In Fig.~\ref{figS3}\textbf{b}, since much more data points are scanned, we use a stronger excitation $\eta_k\bar{\Omega}T\approx 3$ to allow smaller number of repetitions, and just qualitatively demonstrate the suppression of the phonon number after EIT cooling and sideband cooling. In Fig.~\ref{figS3}\textbf{c} we use the same excitation rate $\eta_k\bar{\Omega}T\approx \pi/2$ as in \textbf{a}, and repeat each data point after sideband cooling for 300 times. We then sum over this $150\,$kHz range as an estimation of the average phonon number of all these modes to further suppress the statistical error. Then by similarly comparing the red and the blue motional sidebands, we obtain an average phonon number of $\bar{n}=0.8\pm 0.6$ for the phonon modes relevant for our quantum simulation experiment.

\section{Crystal configuration stability}
\label{sec:stability}
As discussed in Ref.~\cite{PRXQuantum.4.020317}, for a 2D crystal at room temperature, the melting and configuration changes due to collision with background gas molecules are important limiting factors for scaling up the crystal size. Here we show that, for a cryogenic ion trap, such collisions are much less detrimental and the stability of the ion crystal can be significantly improved.

At our temperature $T=6.1\,$K, the residual background gas will mainly consist of hydrogen molecules \cite{Pagano_2019} with a suppressed pressure. To estimate the effect of collision with background gas molecules, we assume head-on elastic collisions as the worst case which transfers the most energy to the collided ion. Suppose the molecule with mass $m$ has a velocity $v$ following a Maxwell-Boltzmann distribution at the temperature $T$ before the collision. After a head-on elastic collision, the collided ion with mass $M$ acquires a velocity $2v/(1+M/m)$. For ${}^{171}\mathrm{Yb}^+$ ions, this velocity is typically an order of magnitude larger than the Doppler temperature, so we can neglect the motion of the ions before the collision. For simplicity, here we consider a 2D ion crystal in a harmonic trap $(\omega_x,\omega_y,\omega_z)=2\pi\times(0.60,2.164,0.144)\,$MHz without micromotion, and postpone the effect of the micromotion later.

\begin{figure}[htbp]
   \includegraphics[width=\linewidth]{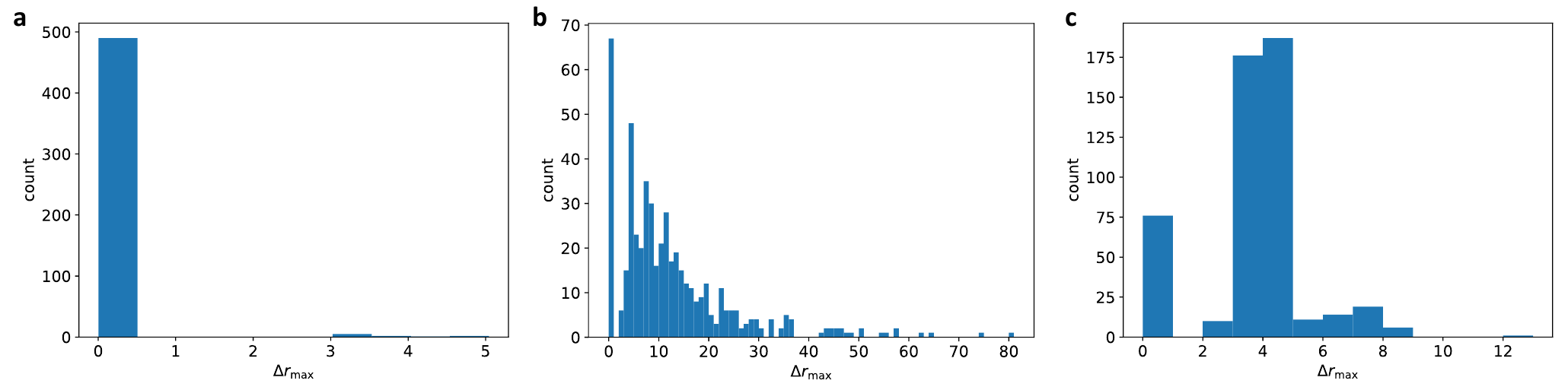}
   \caption{\textbf{a}, Histogram for the maximal position deviation $\max_i \|\vec{r}_i-\vec{r}_i^\prime\|$ after a random head-on elastic collision with a background hydrogen molecule at $T=6.1\,$K and $t=500\,\mu$s Doppler cooling. \textbf{b}, Histogram for the maximal position deviation $\max_i \|\vec{r}_i-\vec{r}_i^\prime\|$ after a random head-on elastic collision with a background hydrogen molecule at $T=300\,$K and $t=500\,\mu$s Doppler cooling. \textbf{c}, Histogram for the maximal position deviation $\max_i \|\vec{r}_{\pi_i}-\vec{r}_i^\prime\|$ at $T=300\,$K after the optimal mapping of ion indices, where $\{\pi_i\}$ is an optimal permutation to minimize the position deviation. Each plot is generated from $500$ random samples.
   \label{fig:collision}}
\end{figure}

We start from an arbitrarily selected equilibrium configuration of $N=512$ ions which is computed by their time evolution under a damping force. After the collision, we assign the velocity to a randomly selected ion, and we numerically simulate the dynamics of all the ions under the harmonic trapping potential, their Coulomb repulsion and a weak damping force from the global Doppler cooling laser. We choose a typical cooling rate $\gamma\approx 8\times 10^3\,\mathrm{s}^{-1}$ for a cooling laser with a saturation parameter $s=1$ and detuning $\Delta=-\Gamma/2$ at equal angle to the three principal axes \cite{RevModPhys.75.281}. As we show in Fig.~\ref{fig:collision}\textbf{a}, after an evolution time of $t=500\,\mu$s which is longer than the relaxation time under the cooling laser, with high probability the ions will remain in their original equilibrium positions. This means that no melting, configuration change or even ion hopping has occurred. Only $2\%$ ($10$ out of $500$) randomly generated samples give a maximal position deviation $\max_i \|\vec{r}_i-\vec{r}_i^\prime\|$ greater than $0.5\,\mu$m, which corresponds to a change in the ion configuration. In other words, the collision of the ion crystal with background gas molecules will hardly change the crystal configuration as long as the Doppler cooling is turned on to dissipate the energy away before it accumulates from multiple collisions. Besides, the cryogenic trap also largely reduces the pressure of the background gas and thus lowering the probability for such collisions to occur.

In comparison, if we set the temperature of the hydrogen molecule to $T=300\,$K, as shown in Fig.~\ref{fig:collision}\textbf{b}, the final positions of the ions typically deviate significantly from their initial positions ($433$ out of $500$ above $1\,\mu$m). We can further show that this deviation comes from both the configuration change of the ions and the global hopping of the ions within the same configuration. As shown in Fig.~\ref{fig:collision}\textbf{c}, we can apply the Hungarian algorithm to find the best mapping between the initial and the final ion indices to minimize their position deviation. Even after this mapping, typically we can find significant change between the initial and final positions ($424$ out of $500$ above $1\,\mu$m), which suggests that the ions have entered a new equilibrium configuration after collision with room-temperature hydrogen molecules.

As for the effect of micromotion, we show in Ref.~\cite{wu2019thesis} that even under micromotion, a 2D ion crystal can still have all its normal mode frequencies to be real, which corresponds to a local minimum in the potential of ions in a harmonic trap. In particular, for the 2D crystals we are considering here, the micromotion amplitudes are still much smaller than the inter-ion spacings, so that the deviation in the ions' equilibrium configuration is also small from that in a harmonic trap. Now given the real mode frequencies, any small perturbation from the equilibrium configuration will be stable and the ideal RF potential will not lead to heating, at least to the lowest order. Even if there may exist higher order heating effects, they can be suppressed by an arbitrarily weak cooling term such as a global Doppler cooling beam. Actually, it is known that RF heating is not significant for ions in the crystal phase, and that it only becomes severe when the ions already acquire sufficient energy to melt into a cloud phase and when their motion becomes chaotic \cite{wu2019thesis}. Therefore, the above analysis that the ion crystal is stable after individual collisions also suggests that RF heating can be neglected in this process.

In the experiment, the stability of our ion crystal is consistent with the above theoretical analysis. The crystal of hundreds of ions hardly melts under global Doppler cooling laser, and we attribute the occasional melting events (once a few days) to the unlocking of the laser frequency. Also we measure a typical crystal configuration lifetime of a few minutes. In comparison, the collision rate between an ion and the background hydrogen molecules at a temperature of $6\,$K and a pressure of $10^{-13}\,$Torr is about $\gamma_{\mathrm{elastic}}\approx 0.01\,\mathrm{s}^{-1}$ \cite{wineland1998experimental,zhang2015thesis}, which corresponds to a timescale below seconds for $N=512$ ions. Note that for a cryogenic trap the pressure is often too low to be measured reliably, and here we are simply using the results in Ref.~\cite{Pagano_2019} as an order-of-magnitude estimation. The actual pressure can be higher, which will then lead to a higher collision rate and thus further support our theoretical analysis that individual collision events will not affect the crystal stability.

\begin{figure}[htbp]
   \includegraphics[width=0.6\linewidth]{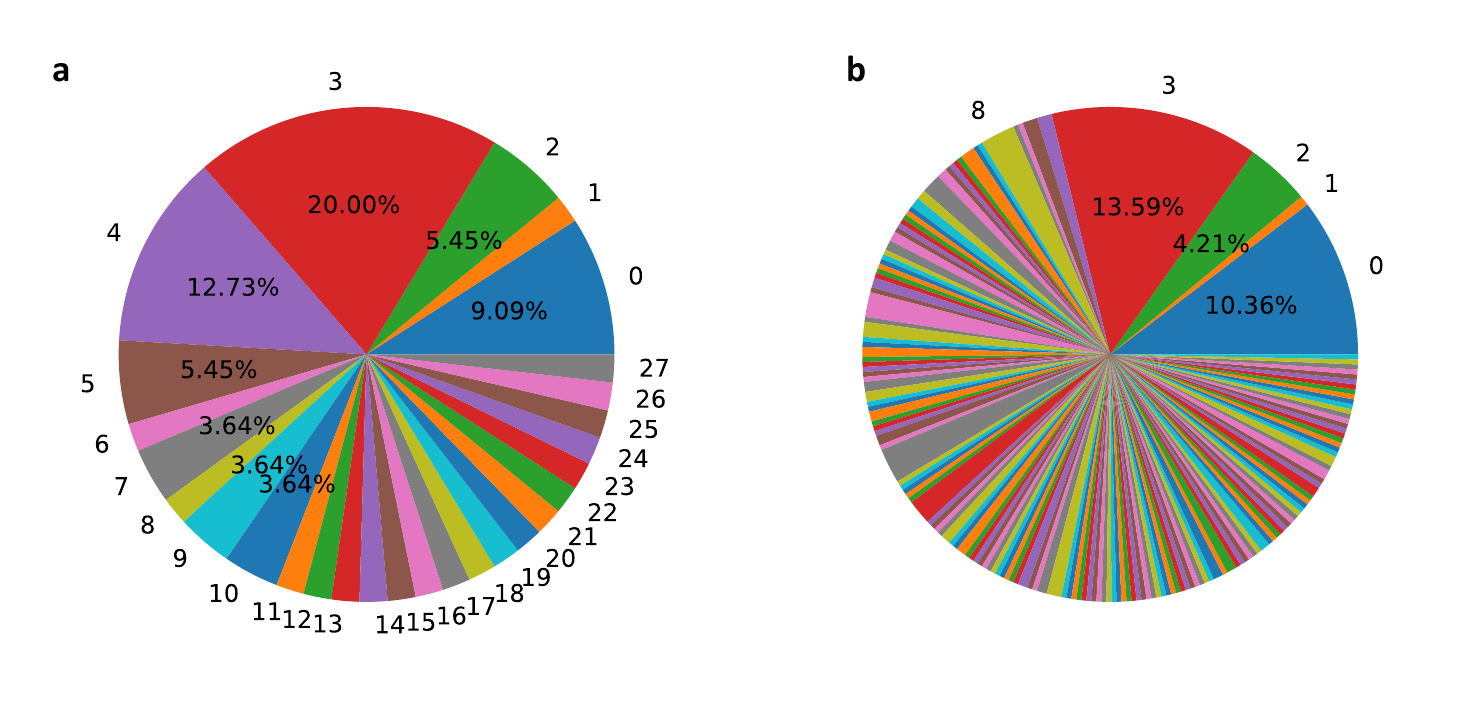}
   \caption{\textbf{a}, Typical pie chart for the statistics of different crystal configurations following a fixed path from the melted state. Different crystal configurations are identified from their image on the CMOS camera. The probabilities are estimated from 55 repetitions. \textbf{b}, After the drift of trap parameters over months and their recalibration, the statistics of different crystal configurations changes, but still we can have sufficient probability to get the desired configuration. In the two plots the same label $0$-$8$ indicates the same crystal configuration.
   \label{fig:configurations}}
\end{figure}

Given the long lifetime of the crystal configuration, we do not require the extremely stable “main configuration” as discussed in Ref.~\cite{PRXQuantum.4.020317}. Instead, after a configuration change is detected, we simply disturb the trap potential to purposely melt the crystal and then slowly recover the trap parameters back. This fixed procedure can result in different crystal configurations, and we can repeat it for multiple times to get the statistics for the probability distribution of these configurations. Then we select the configuration with the highest probability, say, $20\%$ as shown in Fig.~\ref{fig:configurations}, as the targeted one used in the experiment. In this way, we obtain a typical time of about $10\,$s to recover the targeted configuration, which is much shorter than the typical configuration lifetime of a few minutes.
\section{Hamiltonian generated by 411 nm laser}
\label{sec:411}
\subsection{Spin-dependent force}
\label{sec:SDF}
As shown in Fig.~1\textbf{b} of the main text, we use two pairs of counter-propagating $411\,$nm laser to generate the spin-dependent force and further to obtain the spin-spin interaction. For simplicity, we first consider one pair of counter-propagating $411\,$nm beams at the detuning of $\Delta\pm \mu/2$. With a large detuning $|\Delta\pm \mu/2|\gg \Omega$ compared with the Rabi frequency of each laser beam, it is well-known that the beat note of these two beams creates an AC Stark shift varying in time and space, which further leads to a spin-dependent force \cite{leibfried2003experimental,Lee_2005}
\begin{equation}
H=\sum_i\frac{\Omega_i^2}{2\Delta}\left\{1 + \cos\left[\Delta k \cdot y_i(t) - \mu t\right]\right\} |0\rangle_i\langle 0|,
\end{equation}
where $\Delta k$ is the wave vector difference between the two beams along the $y$ direction, and the index $i$ labels all the ions driven by the laser beams. Here the laser only couples $|0\rangle\equiv|^{2}S_{1/2},F=0,m_F=0\rangle$ to the $D_{5/2}$ levels, so that the AC Stark shift only appears for $|0\rangle$ but not $|1\rangle$. The Rabi frequency $\Omega_i$ characterizes the possible spatial variation of the laser intensity.

The second term in the above Hamiltonian gives us the desired spin-dependent force, but the first term can lead to a large longitudinal field which prevents us from observing strong spin-spin correlation in the quantum simulation experiment. Therefore, we further introduce two frequency components at the detuning of $-(\Delta\pm \mu/2)$, so that the overall time-independent AC Stark shifts of the four components cancel with each other. Note that for the second term in the Hamiltonian to add up constructively, we want a relative phase shift of $\pi$ between the two pairs $(\varphi_{b_1}-\varphi_{r_1})-(\varphi_{b_2}-\varphi_{r_2})=\pi$ where $b_1$, $b_2$, $r_1$ and $r_2$ are the four frequency components in Fig.~1\textbf{b} of the main text.

Finally we obtain the Hamiltonian from the four symmetric frequency components
\begin{equation}
H=\sum_i\frac{\Omega_i^2}{\Delta}\cos\left[\Delta k \cdot y_i(t) - \mu t\right] |0\rangle_i\langle 0|.
\end{equation}
When $\mu$ is close to the phonon sideband, we can neglect the far off-resonant terms and get
\begin{equation}
H_{\mathrm{SDF}}=\sum_{ik}\frac{\Omega_i^2}{2\Delta}\left[i \eta_k b_{ik} a_k^\dag e^{-i(\mu-\omega_k)t} + h.c.\right] |0\rangle_i\langle 0|,
\end{equation}
where $\eta_k$ is the Lamb-Dicke parameter and $b_{ik}$ the mode vector. We have neglected higher order expansions of $\eta_k$.
\subsection{Ising model Hamiltonian}
\label{sec:Ising}
The above spin-dependent force differs with the commonly used form \cite{RevModPhys.93.025001} in that it only appears for $|0\rangle$ but not for $|1\rangle$. Therefore, following similar derivation to eliminate the phonon modes adiabatically, we get
\begin{align}
H =& \sum_{ij} \frac{\Omega^i_{\mathrm{eff}} \Omega^j_{\mathrm{eff}}}{4}\sum_k \frac{\eta_k^2 b_{ik} b_{jk}}{\mu-\omega_k} (|0\rangle_i\langle 0|) (|0\rangle_j\langle 0|) \nonumber\\
=&  \sum_{ij} \frac{\Omega^i_{\mathrm{eff}} \Omega^j_{\mathrm{eff}}}{4}\sum_k \frac{\eta_k^2 b_{ik} b_{jk}}{\mu-\omega_k} \frac{I+\sigma_z^i}{2} \frac{I+\sigma_z^j}{2} \nonumber\\
\equiv & \sum_{ij} J_{ij} (I+\sigma_z^i) (I+\sigma_z^j), \label{eq:general}
\end{align}
where $\Omega^i_{\mathrm{eff}}\equiv\Omega_i^2/\Delta$ is the AC Stark shift on the ion $i$ which we deonte as $\Omega_i$ in the main text, and the coupling coefficients $J_{ij}$ are given by
\begin{equation}
J_{ij} = \frac{\Omega^i_{\mathrm{eff}} \Omega^j_{\mathrm{eff}}}{16}\sum_k \frac{\eta_k^2 b_{ik} b_{jk}}{\mu-\omega_k}. \label{eq:Jij}
\end{equation}
In particular, when coupling dominantly to a single mode $k$, we have
\begin{align}
H^{(k)} =& \sum_{ij}\frac{\Omega^i_{\mathrm{eff}} \Omega^j_{\mathrm{eff}}}{16} \frac{\eta_k^2 b_{ik} b_{jk} (I+\sigma_z^i) (I+\sigma_z^j)}{\mu-\omega_k} \nonumber\\
=& \frac{1}{16(\mu-\omega_k)}\left[\sum_i \eta_k b_{ik} \Omega^i_{\mathrm{eff}} (I+\sigma_z^i)\right]^2. \label{eq:single_mode}
\end{align}

Finally, we can bring the spin-spin interaction Hamiltonian in Eq.~(\ref{eq:general}) into the standard form of a quantum Ising model
\begin{equation}
H_{\mathrm{Ising}} = \sum_{i\ne j} J_{ij} \sigma_z^i \sigma_z^j + \sum_i h_i \sigma_z^i,
\end{equation}
where the longitudinal field is given by $h_i \equiv 2\sum_j J_{ij}$. Assuming a nearly constant $\Omega^i_{\mathrm{eff}}$, we have $h_i\propto \sum_j b_{jk}$, which is equal to zero for all the modes apart from the COM mode. As for the COM mode, we can compensate this nearly constant longitudinal field by a small shift in the detuning of the two pairs of the $411\,$nm laser beams, so that a small asymmetry in their time-independent AC Stark shift can be generated. Similarly, in Eq.~(\ref{eq:single_mode}) we can remove the identity matrix $I$ to obtain the Hamiltonian in Eq.~(2) of the main text.


\subsection{Error analysis}
For the ideal Hamiltonian in Eq.~(2) of the main text and the ideal adiabatic evolution, the final state should be an equal superposition of two spin configurations following the pattern of the coupled phonon mode $k$ $\{\sigma_z^i=\mathrm{sign}(b_{ik})\}$ and $\{\sigma_z^i=-\mathrm{sign}(b_{ik})\}$. In particular, when coupled dominantly to the COM mode, the ideal final state should be a GHZ state for $N=300$ qubits. However, in practice various error sources can degrade the fidelity for this macroscopic entangled state. Therefore we do not expect to observe this multipartite entanglement in the current experiment, and leave it as future research directions. Instead, here we mainly focus on the spin-spin correlation which does not require the phase coherence between the two ground states. Below we analyze the dominant error sources in the experiment and explain why we are able to measure a correlation pattern consistent with the phonon modes despite all these errors.

Small uncompensated single-qubit $\sigma_z$ terms can cause an energy bias between the two spin configurations, resulting in an unequal probability to get each of them in the measurement. As we describe above, this effect is the most severe when coupled dominantly to the COM mode, because for the other modes the longitudinal field will largely vanish due to the pattern of the phonon modes. In the experiment, we use this probability distribution, or equivalently the average magnetization of the final state, to calibrate the required AC Stark shift to compensate its effect. This allows us to suppress the residual longitudinal field to be smaller than the Kac normalized Ising coupling. However, note that even $1\,$Hz fluctuation in the longitudinal field can accumulate into considerable phase fluctuation for $N=300$ qubits and an evolution time of about $5\,$ms. Therefore the phase coherence for the two spin configurations in the GHZ state will completely be lost at the current precision. On the other hand, when coupled to other phonon modes, we expect the effect of a global detuning to be less severe, but any asymmetric energy fluctuation between the two ground state spin configurations can still lead to their dephasing.

When simulating a spin model Hamiltonian, the ion trap system is not sensitive to a small phonon number, say, an average thermal phonon number below one, as long as the Lamb-Dicke approximation is still valid \cite{RevModPhys.93.025001}. However, this effective spin-spin coupled model originates from the spin-phonon interaction by adiabatically eliminating the phonon state, so small residual spin-phonon entanglement can still lead to error in the final state. As a toy model, we consider two spins coupled off-resonantly to the COM phonon mode by a spin-dependent force with detuning $\delta$. Under rotating-wave approximation, we have the Hamiltonian
\begin{equation}
H = -\delta a^\dag a + \frac{1}{\sqrt{2}}\eta\Omega (\sigma_z^1 + \sigma_z^2)(a+a^\dag),
\end{equation}
where the factor of $1/\sqrt{2}$ comes from the phonon mode vector. After adiabatically eliminating the phonon mode, we get the standard Ising coupling $H_{\mathrm{eff}}=(\eta^2\Omega^2/\delta)\sigma_z^1 \sigma_z^2$. Now if we start from $|++\rangle$ and adiabatically turn up the spin-spin coupling while adiabatically turn down a transverse field $B(\sigma_x^1+\sigma_x^2)$, we should end up in an EPR state $(|00\rangle+|11\rangle)/\sqrt{2}$.

Now if we consider the spin-phonon coupled Hamiltonian directly and perform the same adiabatic evolution (for simplicity we assume an initial vacuum state for the phonon mode), we expect to end up in the state $(|00\rangle {|\alpha=\sqrt{2}\eta\Omega/\delta\rangle} +|11\rangle {|\alpha=-\sqrt{2}\eta\Omega/\delta\rangle})/\sqrt{2}$. We can trace out the phonon state to obtain the reduced density matrix for the spins, which is effectively a dephasing channel with the off-diagonal term decaying as $\langle -\alpha|\alpha\rangle=\exp(-2|\alpha|^2)$. We can generalize this analysis to $N$ spins (note that we have a mode coefficient proportional to $1/\sqrt{N}$) and get a dephasing term of $\exp(-2 N \eta^2\Omega^2/\delta^2)$, or individual dephasing of $\exp(-2\eta^2\Omega^2/\delta^2)$ for each spin. Again, this error tends to destroy the phase coherence for the ideal GHZ state, but does not directly affect the spin-spin correlation that we measure in the experiment.

\begin{figure}[htbp]
   \includegraphics[width=0.8\linewidth]{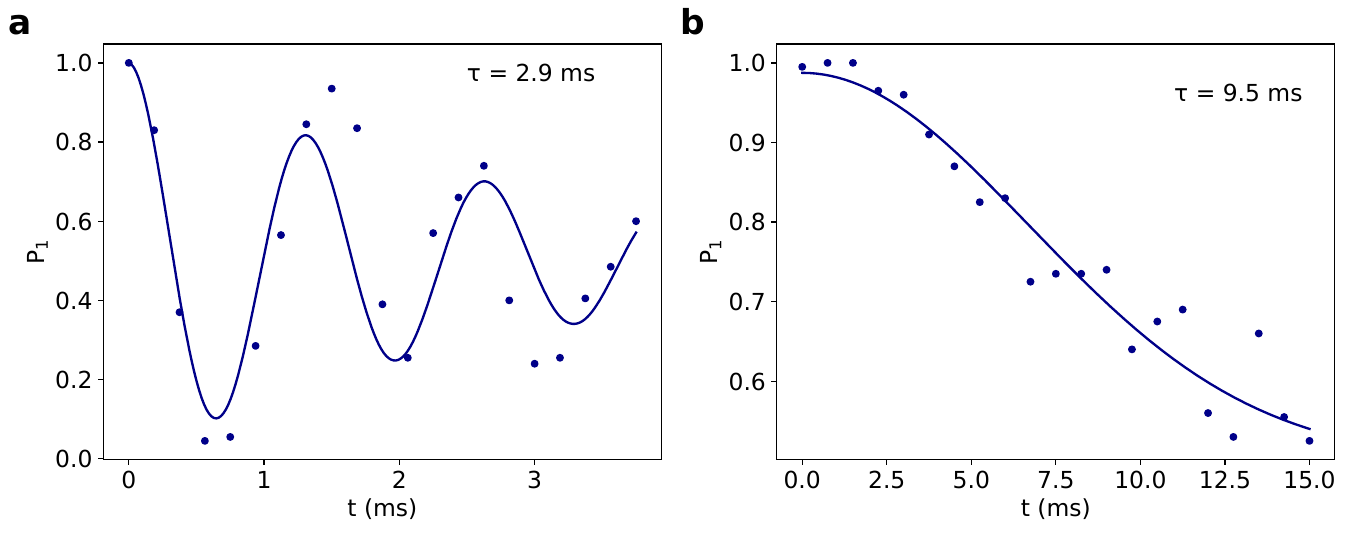}
   \caption{Single-spin coherence time measured by a Ramsey experiment (\textbf{a}) without or  (\textbf{b}) with spin echo when the $411\,$nm laser is turned on.
   \label{fig:coherence}}
\end{figure}
As shown in Fig.~1\textbf{b} of the main text, we use two pairs of symmetric frequency components to cancel their time-independent AC Stark shift on the $|0\rangle$ state, thus suppressing the qubit dephasing due to the laser intensity fluctuation. In practice this cancellation is not perfect and we measure a spin coherence time of about $2.9\,$ms ($9.5\,$ms) without (with) spin echo as shown in Fig.~\ref{fig:coherence} with the $411\,$nm laser turned on. In this calibration we use a laser power comparable to that used for generating the Ising model Hamiltonian, but set the detuning farther away from the phonon modes to avoid phonon excitation. Therefore the measured coherence time shall represent the situation in the quantum simulation experiment. (Note that the low-frequency noise corresponds to an uncompensated $\sigma_z$ term described above, so we expect the coherence time under spin echo to be relevant for the following discussion.) To the lowest order, we can regard this dephasing as time-varying random single-qubit $\sigma_z$ gates on individual ions. Since such terms do not commute with the transverse field during the adiabatic evolution, they will lead to bit flip errors in our final measurement, and thus reduce the spin-spin correlation. Similarly, small state detection error due to imperfect shelving pulses can flip the measured qubit state and lead to the reduction of the correlation.

Finally, there can be non-adiabatic excitation during the slow ramping process. In principle, we can test the adiabaticity by reversing the evolution path and check if the final state returns to the initial one. However, due to the error sources discussed above, we decide that even if the adiabatic condition is satisfied perfectly, the fidelity between the initial and the reversed states will still be vanishingly low. On the other hand, for such non-adiabatic excitations we mainly expect random uncorrelated bit flips in the spin configuration, rather than the structured ones such as the ground state following the pattern of the phonon modes. Therefore we expect that the main effect will again be a decrease in the spin-spin correlation without affecting the measured pattern.
\subsection{Quench parameters}
To prepare the ground state of the quantum Ising model, we apply an initial transverse field $B_0> 50 J_0$ where $J_0\equiv\frac{1}{N}\sum_{i\ne j} J_{ij}$ is the Kac normalized coupling strength, and we follow an exponential path $B(t)=B_0e^{-t/\tau}$ with a total evolution time $T>5\tau$. Specifically, for the data in Fig.~2\textbf{a} of the main text, we set $B_0=2\pi\times 23\,$kHz, $\tau=630\,\mu$s and $T=3.4\,$ms. For the data in Fig.~2\textbf{c} (3\textbf{a}) of the main text, we set $B_0=2\pi\times 8.9\,$kHz, $\tau=1.0\,$ms and $T=5.1\,$ms. For the data in Fig.~2\textbf{e} and \textbf{f} of the main text, we set $B_0=2\pi\times 4.4\,$kHz, $\tau=1.0\,$ms and $T=5.1\,$ms. For the data in Fig.~3\textbf{b} of the main text, we set $B_0=2\pi\times 8.9\,$kHz, $\tau=0.9\,$ms and $T=5.5\,$ms. For the other data in Fig.~3\textbf{c}-\textbf{i} of the main text, we set $B_0=2\pi\times 8.9\,$kHz, $\tau=1.2\,$ms and $T=6.1\,$ms. The results are not sensitive to small changes in these parameters.

\section{Comparison with classical simulated annealing algorithm}
In Fig.~3 of the main text, we measure the spin-spin correlation in the quasi-adiabatically prepared ground states for various Ising models with/without frustration. Here we compare these results with a commonly used classical algorithm, simulated annealing, to verify that our quantum algorithm has achieved reasonable results for most of the cases, and to argue that our output for the most frustrated case is challenging for classical computers.

To obtain the theoretical Ising model Hamiltonian, we first solve the equilibrium configurations of $N=300$ ions in a trap with trap frequencies $(\omega_x,\omega_y,\omega_z)=2\pi\times(0.69,2.140,0.167)\,$MHz. We sort the ions by their $z$ coordinates to match the ion indices used in the experiment. Then we solve all the transverse phonon modes. Note that in general the mode frequencies we get are different from those measured in the experiment due to the anharmonic correction to the trap potential. However, since in the experiment we are dominantly coupling to just one or two phonon modes, we can expect that the computed $J_{ij}$ coefficients or at least its spatial patterns can still agree well with the experimental values, as long as we set the detuning to these modes to be the same as the experiments. Then we can use Eq.~(\ref{eq:Jij}) to compute the theoretical $J_{ij}$ coefficients. Specifically, we use a Rabi rate $\Omega_{\mathrm{eff}}=2\pi\times 10\,$kHz and a Lamb-Dicke parameter of $\eta\approx 0.11$ for the COM mode and counter-propagating $411\,$nm laser beams. Note that for the classical calculation, the overall scaling of the $J_{ij}$ coefficients are not important as it can be absorbed into the inverse temperature for simulated annealing. When two frequency components are applied, we add up their corresponding $J_{ij}$ coefficients together.

Then we solve the ground state for the classical Ising model $H=-\sum_{ij}J_{ij}Z_i Z_j$ (the highest excited state for $H=\sum_{ij}J_{ij}Z_i Z_j$) by the simulated annealing algorithm. Here we set the longitudinal field to zero following the arguments in Sec.~\ref{sec:Ising}. To get the best match to the experimental results, we choose the parameters of the simulated annealing algorithm as $n_{\mathrm{sweep}}=100$ sweeps (each sweep corresponds to $N=300$ attempts to flip a random spin), an initial inverse temperature of $\beta_0=0.01\,\mathrm{kHz}^{-1}$ and a final inverse temperature of $\beta_1=1\,\mathrm{kHz}^{-1}$, and we adjust the inverse temperature linearly from $\beta_0$ to $\beta_1$ within these $n_{\mathrm{sweep}}$ sweeps. We start from a random spin configuration, and update it through the single-spin-flip metropolis algorithm following the above sequence of inverse temperatures. We store the final spin configuration, and repeat this process for $M=100$ times to compute the spin-spin correlation between any pairs. Note that theoretically we already know that this ideal model is symmetric under the global flip of all the spins, so that we can set $\langle Z_i\rangle=0$ for all the spins and simplify $\langle Z_i Z_j\rangle - \langle Z_i\rangle \langle Z_j\rangle = \langle Z_i Z_j\rangle \approx \frac{1}{M}\sum_{k=1}^M Z_i^{(k)} Z_j^{(k)}$.

\begin{figure}[htbp]
   \includegraphics[width=\linewidth]{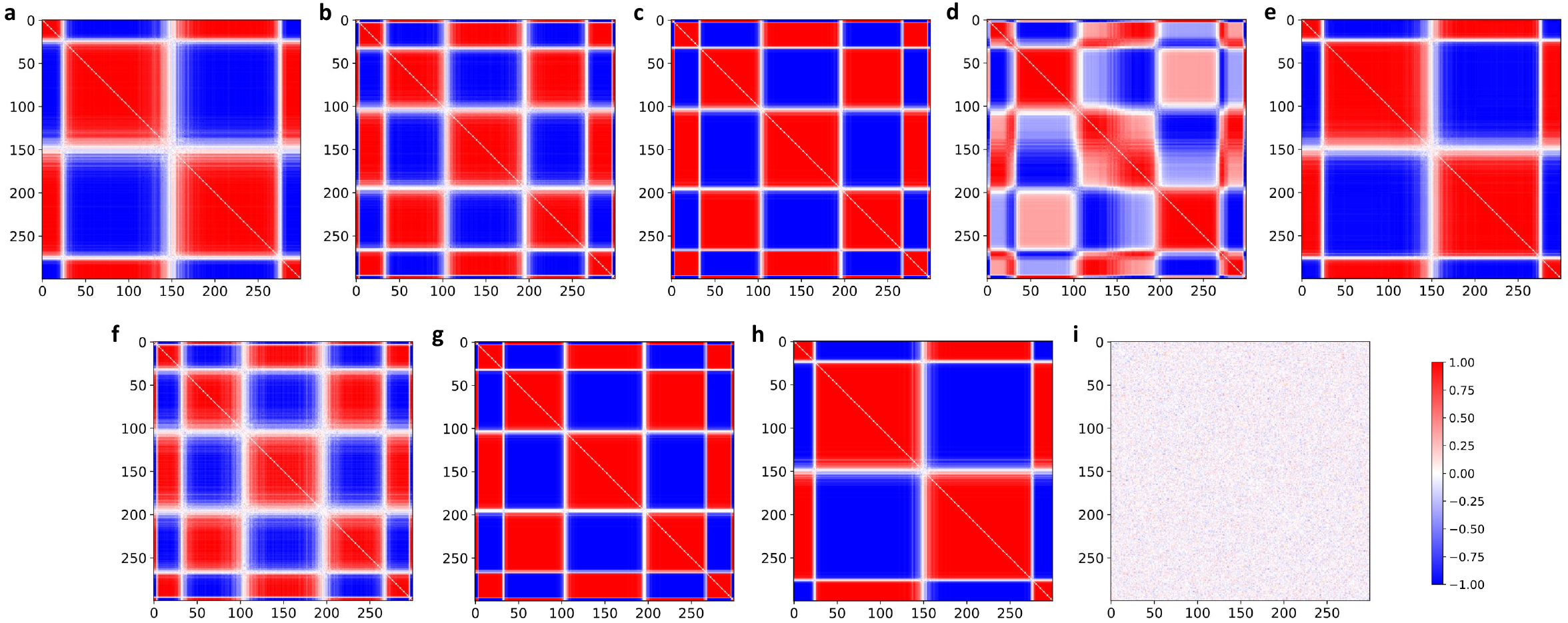}
   \caption{Theoretical results using simulated annealing corresponding to the experimental parameters in Fig.~3 of the main text.
   \label{fig:simulated_annealing}}
\end{figure}

We present the theoretical results in Fig.~\ref{fig:simulated_annealing}. Apart from the most frustrated case in Fig.~\ref{fig:simulated_annealing}\textbf{i}, the rest patterns can be explained by the theoretical ground states well. Note that the classical simulated annealing algorithm and our quantum algorithm are solving the ground states almost independently, with only a few shared parameters like the ion number, the measured trap frequencies and the chosen detuning to the selected phonon modes. Therefore, their agreement suggests that our quantum simulator has successfully achieved the desired Hamiltonian, and has been successful in extracting the desired patterns of the ground states.

Finally, note that in Fig.~\ref{fig:simulated_annealing}\textbf{i} we get almost no signal for the spin-spin correlation, and in particular the pattern does not match that in Fig.~3\textbf{i} of the main text. On the one hand, this suggests that for this frustrated Ising model it is more difficult to get the ground state by the classical algorithm and that larger number of sweeps and lower final temperature will be needed. On the other hand, even if we find the true ground state for this classical Ising model, we do not expect the quantum algorithm to give the same result due to the vanishing energy gap and thus the strong non-adiabatic excitation during the ramping. Therefore, in general we will need to directly compute the 300-qubit quantum dynamics to explain the measured spin-spin correlations, which will be challenging for the available classical computers.

\section{Quantum sampling and coarse graining}
In general, it is difficult to simulate the quantum many-body dynamics such as the long-range transverse-field Ising model in Fig.~4 of the main text by a classical computer, and to sample from the probability distribution when measuring the long-time-evolved final state in, e.g., the computational basis. However, due to the experimental noises, the time-evolved state may gradually decohere and the corresponding probability distribution may approach some trivial distributions like a uniform distribution. To prove that the final distribution is not dominated by experimental decoherence, in Fig.~4\textbf{d} of the main text, we compare the samples from three different values of the transverse field and show that with high probability they come from distinct distributions.

We can use a Pearson's $\chi^2$ test to check if the given samples follow an expected distribution. However, in our experiment, the data locates on a $2^N=2^{300}$-dim space, which is much larger than the sample size of, say, $M=10^3$. Therefore, for general distributions, one expects the data points to scatter randomly over the large space with almost no coincidence, and then it is difficult even to distinguish them from a uniform distribution. Besides, due to the classical difficulty to compute the quantum dynamics and the insufficient experimental sample size, it is also impractical to estimate the probability distribution which we want to compare with. To solve these problems, one possibility is to divide the exponentially large space into a small number of subsets, such that there is considerable probability to get samples in each subset \cite{wang2016coarse}. Specifically, we can get a coarse-grained probability distribution for each of the original distribution at different values of the transverse field. Now if we can show that these coarse-grained distributions are distinct from each other, we can also conclude that the original distributions must be distinguishable.

\begin{figure}[htbp]
   \includegraphics[width=0.3\linewidth]{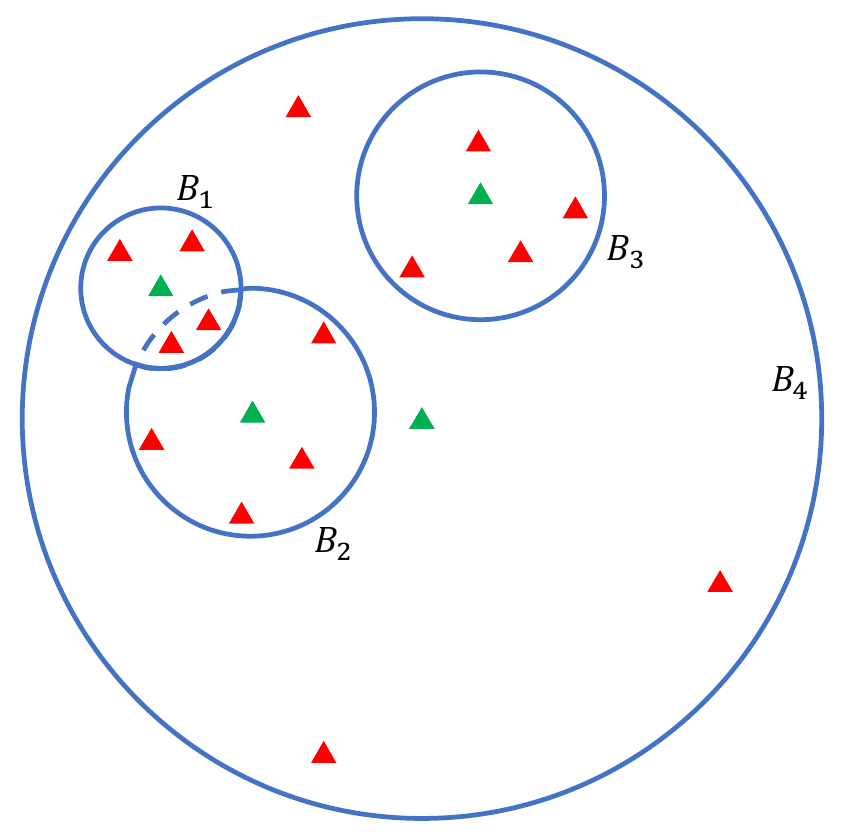}
   \caption{Schematic for the construction of coarse-graining bubbles given a reference sample set. Triangles are the data points in the sample, and the green ones are selected as the centers of bubbles. $B_1$ to $B_4$ indicate the order in constructing the bubbles, and the blue solid curves indicate the boundary between bubbles. Here we require all the bubbles except for the last one to contain at least $m=5$ data points for illustration. In real experiments we use $m=500$. The last bubble covers the whole space.
   \label{fig:bubbles}}
\end{figure}

As shown in Fig.~\ref{fig:bubbles}, we use the following procedure to design the subsets, or ``bubbles'', for coarse graining from a large sample which we call the ``reference''. We choose a target sample size $m$ and require each bubble to contain at least $m$ data points in the reference. We start from a random data point in the reference, and design a subset to be a bubble centered at this point with the smallest possible radius to include no less than $m$ data points in it (with the distance measured by Hamming distance between spin configurations). Then we remove all the data points in the selected bubble from the reference set, and we pick another random data point to construct the next bubble. We repeat this process until the reference set becomes empty, and the last constructed bubble will cover the whole $2^{N}$-dim space. Note that when constructing the bubbles, we record their order. Any given data point will be assigned to the first bubble within this sequence that can cover it. For example, the two data points at the intersection of $B_1$ and $B_2$ in Fig.~\ref{fig:bubbles} will be assigned to $B_1$.

A potential loophole in this construction is that, the centers and radii of the bubbles are determined by the reference and are thus biased. Therefore, they may not correctly represent the underlying probability distribution. As an extreme case, if we choose $m=1$ in the construction, our bubbles will just be all the data points in the reference set with a radius of zero, together with a last bubble to cover the remaining space. Now if we generate a new sample with a similar size, with high probability we will get no counts in these small bubbles at all, with all the data points locating in the last bubble. This will lead to an apparent rejection of samples following the same distribution as the reference, and is not what we want. Fortunately, this problem can be relieved if we use large $m$ and thus smaller number of bubbles and smaller degrees of freedom in their parameters. On the other hand, we do not want $m$ to be too large because in the other extreme case when we have a single bubble covering the whole space, we will not be able to distinguish any probability distributions.

In the experiment, we use a large reference set of $4912$ data points at the transverse-field of $B_0=1.43 J_0$ to construct the bubbles, and we set $m=500$ to give us in total 10 bubbles. The expected probability distribution within these bubbles can thus be estimated by the data counts in each bubble as shown in Fig.~4\textbf{d} of the main text. Then we take another independent sample of $1000$ data points, also at $B_0=1.43 J_0$, to test if the bias in the bubbles is sufficiently low. As we show in the main text, the $\chi^2$ test gives a $p$-value of $0.8$ and supports the null hypothesis that the sample is from the expected distribution. In comparison, when we take $1000$ data points from different transverse fields like $B_0=0.71 J_0$ or $B_0=2.14 J_0$, the histogram looks very different and we get a $p$-value below $10^{-49}$, which strongly rejects the null hypothesis. Therefore, we prove that with high probability our quantum sampling from evolved states under different values of the transverse field gives distinct probability distributions that are not dominated by the experimental decoherence.

Although in general the many-body quantum dynamics is intractable for classical computers, in some special cases efficient solutions may exist. A relevant situation for the long-range transverse-field Ising model we are considering is when the interaction range becomes infinity, namely an all-to-all coupled Hamiltonian
\begin{equation}
H = \frac{J_0}{N - 1} \sum_{i\ne j} \sigma_z^i \sigma_z^j + B_0 \sum_i \sigma_x^i = \frac{J_0}{N - 1} \left(\sum_i \sigma_z^i\right)^2 + B_0 \sum_i \sigma_x^i \equiv \frac{4J_0}{N - 1} J_z^2 + 2B_0 J_x,
\end{equation}
where $J_x$ and $J_z$ are angular momentum operators for a particle with total angular momentum $j=N/2$, and we have dropped an irrelevant constant in the Hamiltonian. Note that our initial polarized state along the $\sigma_z$ direction corresponds to $|j=N/2,m=N/2\rangle$, hence ideally the system will stay in this $(N+1)$-dim Hilbert space during the time evolution, which can be simulated efficiently.
\begin{figure}[htbp]
   \includegraphics[width=0.6\linewidth]{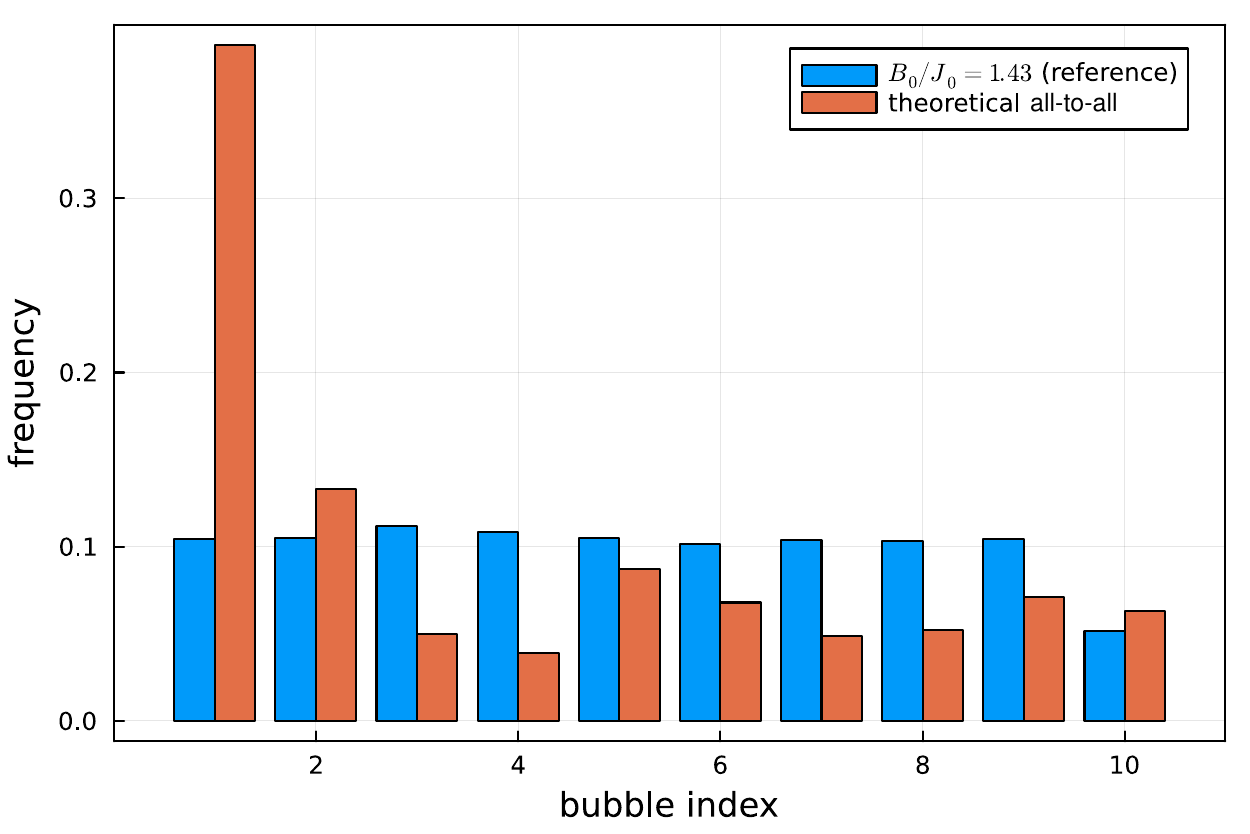}
   \caption{Comparison between the coarse-grained experimental probability distribution (which does not follow exact all-to-all coupling) and that from 1000 random samples generated by the theoretical all-to-all coupled model with the same parameters $J_0=2\pi\times 0.31\,$kHz, $B_0/J_0=1.43$ and an evolution time $T=6\,$ms. The $p$-value is below $10^{-99}$.
   \label{fig:Dicke}}
\end{figure}

In Fig.~\ref{fig:Dicke} we compare this theoretical distribution with the experimentally measured results and find them to be distinct from each other. This is understandable since the actual Ising model we generate in the experiment has contribution from other phonon modes as well, and any such deviation or nonuniformity in our Hamiltonian, or any experimental noise, can break the symmetry in the all-to-all coupled model and lead to an evolution in an exponentially higher dimensional space. Therefore, we conclude that to the best of our knowledge, there is no efficient classical algorithm to generate samples that are indistinguishable with our experimental results.
\bibliography{reference}